\documentclass[journal]{IEEEtran}
\ifCLASSINFOpdf
\else
\fi

\usepackage{graphicx}
\usepackage{amssymb}
\usepackage{amsmath}
\usepackage{booktabs}
\usepackage{multirow}
\usepackage{enumitem}

\graphicspath{{figures/}}
\usepackage[outdir=figures_pdf/]{epstopdf}

\usepackage{xcolor}
\definecolor{ECS-Blue}{rgb}{0,0.4392,0.7529}
\definecolor{ECS-Red}{rgb}{0.7529,0,0}
\definecolor{ECS-Green}{rgb}{0.4667,0.5765,0.2353}

\usepackage[para]{threeparttable}
\makeatletter
\def\TPT@doparanotes{\par
   \prevdepth\z@ \TPT@hsize
   \TPTnoteSettings
   \parindent\z@ \pretolerance 8
   \linepenalty 200
   \renewcommand\item[1][]{\relax\ifhmode \begingroup
       \unskip
       \advance\hsize 10em 
       \penalty -45 \hskip\z@\@plus\hsize \penalty-19
       \hskip .15\hsize \penalty 9999 \hskip-.15\hsize
       \hskip .01\hsize\@plus-\hsize\@minus.01\hsize 
       \hskip 0em\@plus .3em
      \endgroup\fi
      \tnote{##1}\,\ignorespaces}%
   \let\TPToverlap\relax
   \def\endtablenotes{\par}%
}

\begin{document}

\renewcommand{\theenumi}{\alph{enumi}}
\setlength{\abovedisplayskip}{4pt plus 1pt minus 1pt}
\setlength{\belowdisplayskip}{4pt plus 0pt minus 1pt}
\setlength{\abovecaptionskip}{0pt plus 1pt minus 1pt}
\setlength{\belowcaptionskip}{0pt plus 0pt minus 1pt}
\setlength{\textfloatsep}{3pt plus 1pt minus 1pt}
\setlength{\floatsep}{3pt plus 0pt minus 1pt}
\setlength{\dbltextfloatsep}{3pt plus 1pt minus 1pt}
\setlength{\dblfloatsep}{3pt plus 0pt minus 1pt}

%
\title{A 2.5-nA Area-Efficient Temperature-Independent\\
176-$\:$/$\:$82-ppm/$^\circ$C CMOS-Only Current Reference\\
in 0.11-$\mu$m Bulk and 22-nm FD-SOI}
%
%
%

\author{Martin~Lefebvre,~\IEEEmembership{Graduate Student Member,~IEEE},
        and David~Bol,~\IEEEmembership{Senior Member,~IEEE}
\vspace{-0.75cm}
\thanks{Manuscript received 17 November 2023; revised 28 February 2024 and 21 April 2024; accepted 15 May 2024. This article was approved by Associate Editor Fabio Sebastiano. This work was supported by the Fonds de la Recherche Scientifique (FRS-FNRS) of Belgium under grant CDR J.0014.20. \textit{(Corresponding author: Martin Lefebvre.)}

The authors are with the Université catholique de Louvain, Institute of Information and Communication Technologies, Electronics and Applied Mathematics, B-1348 Louvain-la-Neuve, Belgium (e-mail: \{martin.lefebvre; david.bol\}@uclouvain.be).

Color versions of one or more figures in this article are available at https://doi.org/10.1109/JSSC.2024.3402960.

Digital Object Identifier 10.1109/JSSC.2024.3402960
}}

%
%

\markboth{IEEE Journal of Solid-State Circuits,~Vol.~xx, No.~xx, xx~2024}%
{Shell \MakeLowercase{\textit{et al.}}: Bare Demo of IEEEtran.cls for IEEE Journals}
%
\IEEEoverridecommandlockouts
\IEEEpubid{\begin{minipage}{\textwidth}\ \\[12pt] \begin{scriptsize}This document is the paper as accepted for publication in JSSC, the fully edited paper is available at https://ieeexplore.ieee.org/document/10550946. \copyright 2024 IEEE. Personal use of this material is permitted. Permission from IEEE must be obtained for all other uses, in any current or future media, including reprinting/republishing this material for advertising or promotional purposes, creating new collective works, for resale or redistribution to servers or lists, or reuse of any copyrighted component of this work in other works.\end{scriptsize}
\end{minipage}}
\maketitle

\begin{abstract} Internet-of-Things (IoT) applications require \mbox{nW-power} current references that are robust to process, voltage and temperature (PVT) variations, to maintain the performance of IoT sensor nodes in a wide range of operating conditions. However, nA-range current references are rarely area-efficient due to the use of large gate-leakage transistors or resistors, which occupy a significant silicon area at this current level. In this paper, we introduce a nA-range constant-with-temperature (CWT) current reference, relying on a self-cascode \mbox{MOSFET} (SCM) biased by a four-transistor ultra-low-power voltage reference through a single-transistor buffer. The proposed reference includes a temperature coefficient (TC) calibration mechanism to maintain performance across process corners. In addition, as the proposed design relies on the body effect, it has been fabricated and measured in \mbox{0.11-$\mu$m} bulk and \mbox{22-nm} fully-depleted silicon-on-insulator (\mbox{FD-SOI}) to demonstrate feasibility in both technology types. On the one hand, the \mbox{0.11-$\mu$m} design consumes a power of 16.8~nW at 1.2~V and achieves a \mbox{2.3-nA} current with a line sensitivity (LS) of 2.23~$\%$/V at 25$^\circ$C and a TC of 176~ppm/$^\circ$C at 1.2~V from -40 to 85$^\circ$C. On the other hand, the \mbox{22-nm} design consumes a power of 16.3~nW at 1.5~V and achieves a \mbox{2.5-nA} current with a \mbox{1.53-$\%$/V} LS at 25$^\circ$C and an \mbox{82-ppm/$^\circ$C} TC at 1.5~V from -40 to 85$^\circ$C. Thanks to their simple architecture, the proposed references achieve a silicon area of 0.0106~mm$^2$ in 0.11~$\mu$m and 0.0026~mm$^2$ in 22~nm without compromising other figures of merit, and are thus competitive with state-of-the-art CWT references operating in the same current range.
\end{abstract}


%
\IEEEpeerreviewmaketitle

\vspace{-0.25cm}
\section{Introduction}
\label{sec:1_introduction}
%
%
%
%
\begin{figure}[!t]
	\centering
	\includegraphics[width=.49\textwidth]{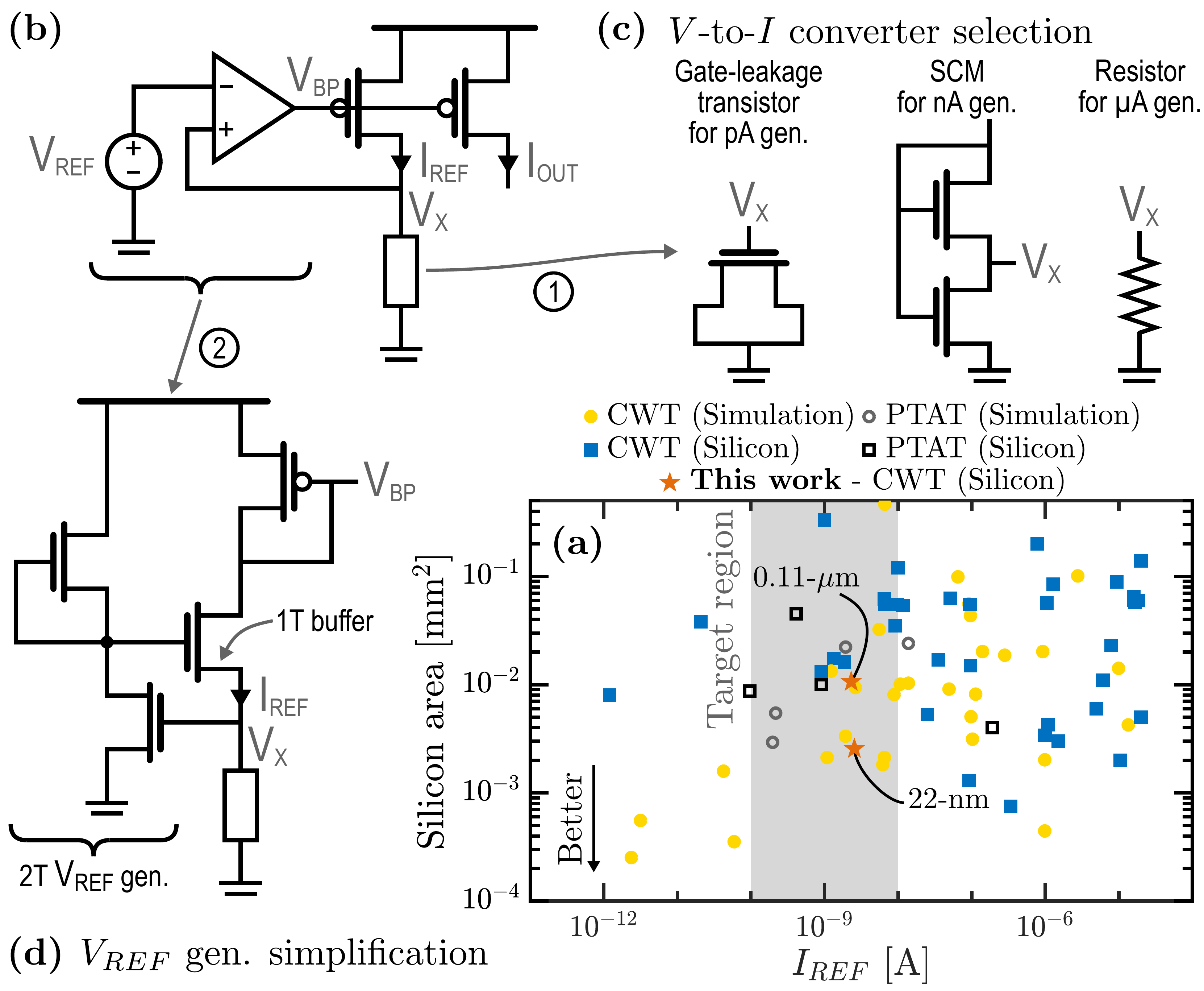}
	\caption{\textbf{The area efficiency of nA-range CWT current references can be improved by using an SCM as $\boldsymbol{V}$-to-$\boldsymbol{I}$ converter, and by simplifying the $\boldsymbol{V_{REF}}$ generation and buffering.} (a) Tradeoff between silicon area and reference current, featuring the scarcity of measured area-efficient solutions in the nA range. (b) Conventional CWT current references are based on a reference voltage applied to a $V$-to-$I$ converter, which can either be (c) a gate-leakage transistor, an SCM, or a resistor, respectively well suited to the generation of pA-, nA-, or $\mu$A-range reference current. (d) Voltage reference implemented with a 2T structure and a 1T buffer \cite{Wang_2018, Lefebvre_2022}.}
	\label{fig:1_context}
\end{figure}
\IEEEPARstart{I}{n} the last decade, the expected growth of the Internet of Things (IoT) has fostered the development of ultra-low-power (ULP) smart sensor nodes, combining sensing and processing capabilities at the edge. As most integrated circuits, the analog blocks constituting these sensor nodes must be biased by current references, whose specifications can be derived from the requirements of IoT applications. First, sensor nodes are usually supplied by limited-capacity batteries or energy harvesting, constraining the average power consumption to the 0.1-to-100-$\mu$W range \cite{Blaauw_2014}. A bias current in the order of nA is thus required to cope with the stringent power constraints of always-on blocks in sleep mode while maintaining a satisfying performance in active mode. Moreover, sensor nodes must be able to operate in a wide range of deployment scenarios. The sensitivity of current references to process, voltage and temperature (PVT) variations should therefore be mitigated, as it can profoundly undermine the performance of common analog building blocks such as real-time clock generators \cite{Wang_2017_ESSCIRC, Liao_2023}, temperature sensors \cite{Jeong_2014, Wang_2017_Nature}, and capacitive sensor interfaces \cite{Omran_2014}. Finally, silicon area must be limited to reduce the production cost and direct environmental footprint of each sensor, especially given the expected massive production volume of the IoT \cite{Pirson_2021}. The same trend should be followed by current references. However, Fig.~\ref{fig:1_context}(a) highlights that nA-range current references that are simultaneously robust to PVT variations and area-efficient are scarce in the literature, or have never been demonstrated in measurement.\\
\IEEEpubidadjcol
\indent One of the conventional ways of generating a reference current is to bias a voltage-to-current (\mbox{$V$-to-$I$}) converter with a reference voltage through an op-amp-based feedback loop [Fig.~\ref{fig:1_context}(b)]. On the one hand, the choice of \mbox{$V$-to-$I$} converter \textcircled{\raisebox{-0.9pt}{1}} strongly impacts the area efficiency of the reference. For example, gate-leakage transistors and resistors are not well suited to the generation of a nA-range current as they occupy a significant area at this current level [Figs.~\ref{fig:1_context}(a) and (c)]. Previous works have shown that self-cascode MOSFETs (SCMs) are more appropriate to generate a nA-range current \cite{CamachoGaleano_2005, CamachoGaleano_2008}, but need to be biased by a proportional-to-absolute-temperature (PTAT) voltage with a constant-with-temperature (CWT) offset to make the current CWT \cite{Lefebvre_2023}. A temperature coefficient (TC) calibration mechanism is also necessary to maintain performance across process corners, but is difficult to integrate in the $\beta$-multiplier architecture proposed in \cite{Lefebvre_2023}. In addition, the value of the CWT offset in \cite{Lefebvre_2023} depends solely on technological parameters and cannot be tuned through transistor sizes. This makes the attainable performance heavily reliant on technology, potentially leading to some degradation when the CWT offset is not adapted. On the other hand, the last decade has seen ULP voltage references composed of only a few transistors \textcircled{\raisebox{-0.9pt}{2}} arising as simple and area-efficient alternatives to more complex architectures, with only a marginal deterioration in robustness \cite{Adriaensen_2002, Seok_2012, CamposDeOliveira_2018, Fassio_2021}. Such voltage references, coupled with a single-transistor (1T) buffer [Fig.~\ref{fig:1_context}(d)], have already been successfully applied to the generation of a wide range of reference currents \cite{Wang_2018, Zhuang_2020, Lefebvre_2022}, but never to a CWT nA-range current.\\
\indent In this work, we propose to fix the two critical issues of \cite{Lefebvre_2023} mentioned hereabove by generating the SCM bias voltage with a four-transistor (4T) voltage reference. This structure includes a calibration mechanism to tune the TC in process corners, makes the CWT offset dependent on transistor sizes, and removes the need for a startup circuit. This new building block is the key innovation of this work, as it allows to reliably attain a good performance in terms of LS, TC, and variability in any technology, while minimizing area and power consumption. This leads to an area-efficient CWT current reference generating a current in the nA range and relying on an SCM biased by this 4T voltage reference through a 1T buffer. Note that the proposed architecture can also be used in applications requiring a bias current with a specific temperature dependence, by using a different sizing of the reference or by leveraging the TC calibration mechanism to configure the TC.
\begin{figure}[!t]
	\centering
	\includegraphics[width=.488\textwidth]{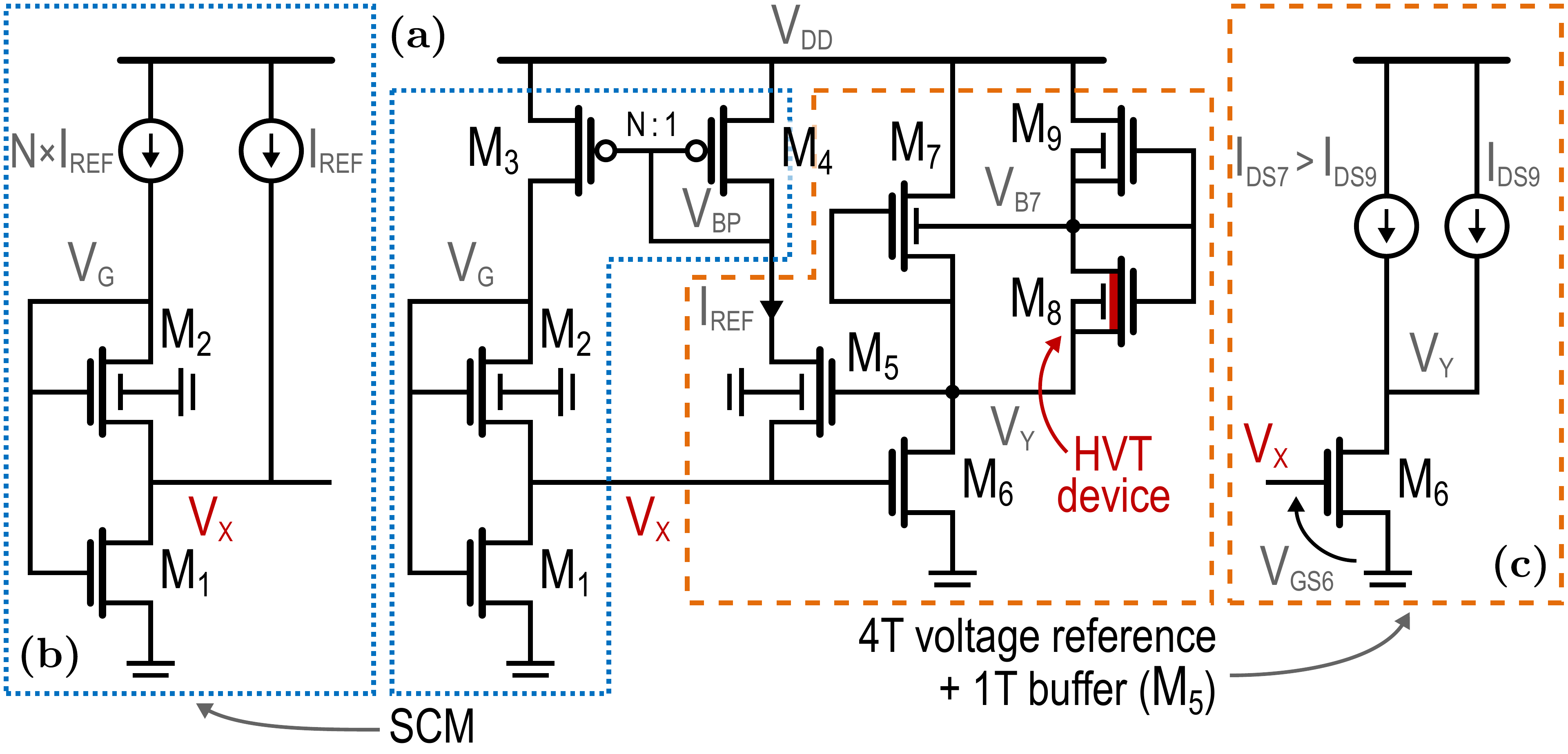}
	\caption{(a) Basic schematic of the proposed current reference, which consists of (b) an SCM ($M_{1-2}$) biased by a pMOS current mirror ($M_{3-4}$), and (c) a 4T voltage reference ($M_{6-9}$) generating a PTAT voltage with a CWT offset, biasing the SCM through a 1T buffer ($M_5$).}
	\label{fig:2_basic_schematic}
\end{figure}
Two prototypes have been fabricated in United Microelectronics Corporation (UMC) \mbox{0.11-$\mu$m} bulk and GlobalFoundries (GF) \mbox{22-nm} fully-depleted silicon-on-insulator (FD-SOI) CMOS technologies. The \mbox{0.11-$\mu$m} (resp. \mbox{22-nm}) design show a measured TC of 176~ppm/$^\circ$C (resp. 82~ppm/$^\circ$C) from \mbox{-40} to 85$^\circ$C, while occupying a silicon area of only 0.0106~mm$^2$ (resp. 0.0026~mm$^2$). The rest of this work is structured as follows. Section~\ref{sec:2_governing_equations_and_operating_principle} details the governing equations of the reference and presents its operation principle. Next, Section~\ref{sec:3_design_and_sizing_methodology} discusses the design and sizing methodology, while Section~\ref{sec:4_simulation_and_measurement_results} examines the simulation and measurement results. Finally, Section~\ref{sec:5_comparison_to_the_state_of_the_art} compares this work to the literature while Section~\ref{sec:6_conclusion} delivers concluding remarks.

\section{Governing Equations and Operating Principle}
\label{sec:2_governing_equations_and_operating_principle}
This section presents the equations governing the behavior of the two constitutive blocks of the proposed current reference, namely an SCM and a 4T voltage reference, depicted in Fig.~\ref{fig:2_basic_schematic}(a) in its basic form, i.e., without calibration circuit. It also details the operation principle of the reference, and explains how the limitations of the SCM-based $\beta$-multiplier current reference in \cite{Lefebvre_2023} have been overcome in Section~\ref{subsec:2C_proposed_temperature_independent_current_reference}.\\
\indent In what follows, the line sensitivity (LS) and TC are computed using the box method, i.e.,
\begin{IEEEeqnarray}{RCL}
	\textrm{LS} & = & \frac{\left(I_{REF,\mathrm{max}}-I_{REF,\mathrm{min}}\right)}{I_{REF,\mathrm{avg}} \left(V_{DD,\mathrm{max}}-V_{DD,\mathrm{min}}\right)} \times 100 \: \%\textrm{/V,}\label{eq:LS}\\
	\textrm{TC} & = & \frac{\left(I_{REF,\mathrm{max}}-I_{REF,\mathrm{min}}\right)}{I_{REF,\mathrm{avg}} \left(T_{\mathrm{max}}-T_{\mathrm{min}}\right)} \times 10^6 \: \textrm{ppm/}^\circ\textrm{C,}\label{eq:TC}%
\end{IEEEeqnarray}
where $I_{REF,\mathrm{min/avg/max}}$ respectively stand for the minimum, average, and maximum reference current among the considered range. $V_{DD,\mathrm{min/max}}$ (resp. $T_{\mathrm{min/max}}$) refer to the lower and upper bounds of the voltage (resp. temperature) range.

\subsection{Self-Cascode MOSFET}
\label{subsec:2A_self_cascode_MOSFET}
\begin{figure}[!t]
	\centering
	\includegraphics[width=.48\textwidth]{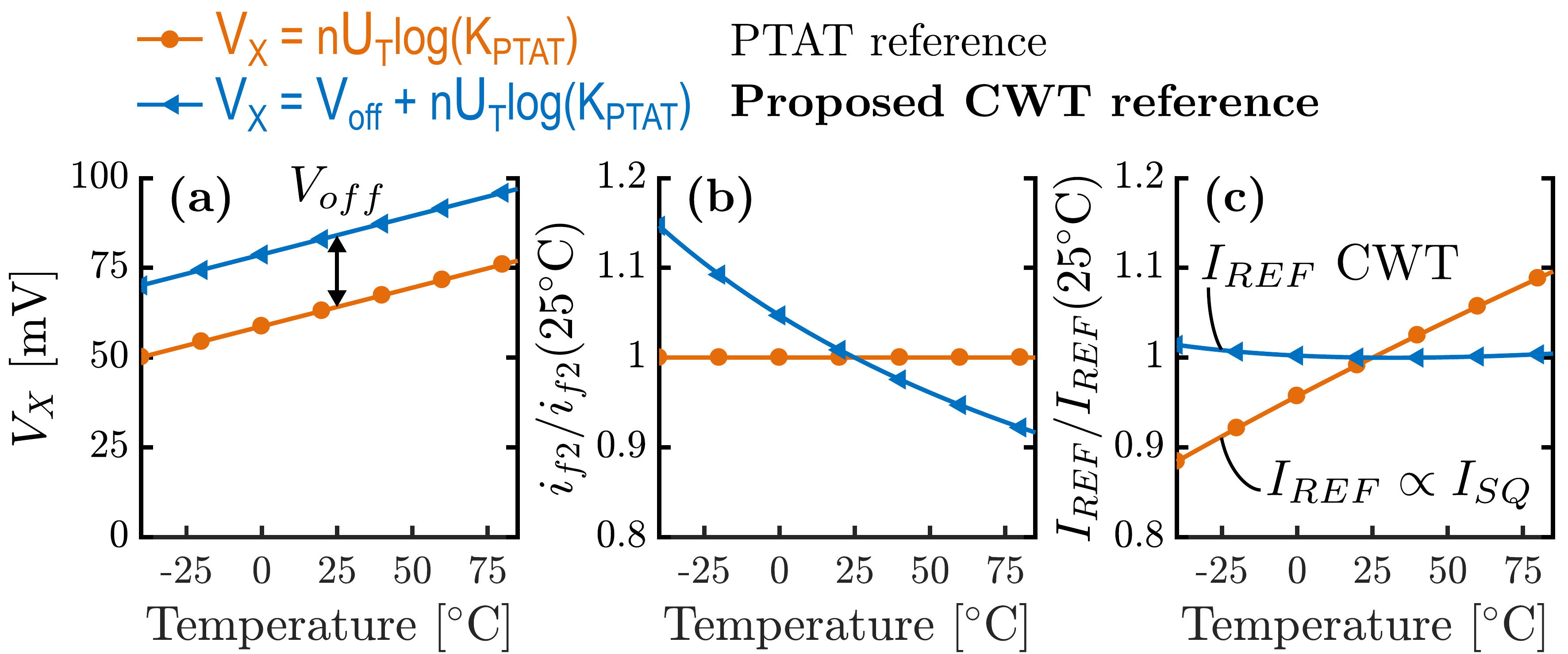}
	\caption{Operation principle of PTAT references proposed in prior art \cite{CamachoGaleano_2005, CamachoGaleano_2008} (in orange) and of the proposed CWT reference (in blue). Analytical expression of (a) the voltage $V_X$ applied to the SCM, (b) the inversion level of $M_2$, denoted as $i_{f2}$, and (c) the reference current $I_{REF}$, as a function of temperature and for $V_{off}$~=~20~mV. Generic technological parameters $n$~=~1.2 and $m$~=~1.5 are selected. (b) and (c) are normalized to their value at 25$^\circ$C. For the proposed CWT reference, the parameters leading to a minimum $I_{REF}$ TC are $K_{PTAT} = 8$ and $\alpha = 1.5$.}
	\label{fig:3_scm_operation_principle}
\end{figure}

First, the SCM depicted in Fig.~\ref{fig:2_basic_schematic}(b) relies on long-channel transistors in moderate inversion, for which a simplified model such as the analog compact MOSFET (ACM) model \cite{Cunha_1998} is adequate to describe the transistor current-voltage (\mbox{$I$-$V$}) characteristics. In this model, the drain current is given by
\begin{equation}
	I_D = I_{SQ}S(i_f-i_r)\textrm{,}\label{eq:id_acm}
\end{equation}
where $I_{SQ} = \frac{1}{2}\mu C_{ox}^{'} nU_T^2$ is the specific sheet current, $\mu$ is the carrier mobility, $C_{ox}^{'}$ is the gate oxide capacitance per unit area, $n$ is the subthreshold slope factor, $U_T$ is the thermal voltage, $S = W/L$ is the transistor aspect ratio, and $i_f$, $i_r$ are the forward and reverse inversion levels. The transistor $I$-$V$ curve is captured by
\begin{equation}
	V_P - V_S = U_T \left[\sqrt{1+i_f} - 2 + \log\left(\sqrt{1+i_f}-1\right)\right]\textrm{,}\label{eq:vs_acm}
\end{equation}
where $V_P = (V_G - V_{T0})/n$ is the pinch-off voltage, $V_{T0}$ is the threshold voltage at zero $V_{BS}$, and all voltages are referred to the transistor's body, thus accounting for the body effect. $i_{r}$ is found by replacing $V_S$ by $V_D$ in (\ref{eq:vs_acm}), but is only relevant when the transistor is not saturated. It should be noted that, in what follows, the obtained expressions only slightly differ from the ones obtained in \cite{Lefebvre_2023} due to the body connection of $M_2$ to ground. We have nonetheless chosen to detail them here to avoid any mistake and to provide all the necessary tools for implementing the proposed reference. Applying the ACM equations to transistors $M_{1-2}$ leads to the following equation, which expresses voltage $V_X$ as
\begin{IEEEeqnarray}{RCL}
	V_X & = & U_T \Bigg[\left(\sqrt{1+\alpha i_{f2}}-\sqrt{1+i_{f2}}\right)\IEEEnonumber\\
	& & + \log\left(\frac{\sqrt{1+\alpha i_{f2}}-1}{\sqrt{1+i_{f2}}-1}\right)\Bigg]\label{eq:vx_SCM_if1}
\end{IEEEeqnarray}
by defining $\alpha \triangleq i_{f1}/i_{f2} > 1$. Then, based on (\ref{eq:id_acm}) applied to $M_{1-2}$, the ratio of the aspect ratios $S_1/S_2$ must comply with
\begin{equation}
	\frac{S_1}{S_2} = \frac{I_{SQ2}}{I_{SQ1}} \frac{1+N}{N} \frac{1}{\alpha - 1}\label{eq:S1_over_S2}
\end{equation}
to ensure that Kirchhoff's current law is respected \cite{CamachoGaleano_2008}, $N$ being the current ratio between $M_3$ and $M_4$. Besides, applying (\ref{eq:id_acm}) to $M_2$ gives the expression of the reference current
\begin{equation}
	I_{REF}(T) = I_{SQ2}(T) i_{f2}(T) (S_2/N)\textrm{,}\label{eq:iref}
\end{equation}
where $I_{SQ2}(T) \propto U_T^2\mu(T) \propto T^{2-m}$, with $\mu(T) = \mu(T_0)\left(T/T_0\right)^{-m}$, and $m$ is the temperature exponent of the carrier mobility whose value is comprised between 1.2 and 2 in bulk CMOS \cite{Tsividis_1999}. An important quantity which remains to be defined is the sensitivity of the reference current to $V_X$, computed with $I_{REF}$ expressed by (\ref{eq:iref}) and $di_{f2}/dV_X$ calculated from (\ref{eq:vx_SCM_if1}) as $(dV_X/di_{f2})^{-1}$, consequently yielding
\begin{equation}
	S_{I_{REF}} = \frac{2}{i_{f2} U_T}\left[\frac{\alpha}{\sqrt{1+\alpha i_{f2}}-1} - \frac{1}{\sqrt{1+i_{f2}}-1}\right]^{-1}\textrm{.}\label{eq:siref}
\end{equation}
In existing SCM-based PTAT current references \cite{CamachoGaleano_2005, CamachoGaleano_2008, Lefebvre_2022}, the SCM is biased by a purely PTAT voltage $V_X$ [Fig.~\ref{fig:3_scm_operation_principle}(a)] and $M_{1-2}$ operate at a fixed inversion level [Fig.~\ref{fig:3_scm_operation_principle}(b)], thus making $I_{REF}$ proportional to the specific sheet current $I_{SQ}$ (\ref{eq:iref}). Therefore, $I_{REF}$ presents a PTAT behavior dictated by $T^{2-m}$ [Fig.~\ref{fig:3_scm_operation_principle}(c)]. Nevertheless, to obtain an SCM-based CWT current reference such as \cite{Lefebvre_2023}, the SCM must be biased by
\begin{equation}
	V_X = V_{off} + nU_T\log\left(K_{PTAT}\right)\textrm{,}\label{eq:vx_generic}
\end{equation}
i.e., a PTAT voltage with a CWT offset as shown in Fig.~\ref{fig:3_scm_operation_principle}(a), with which solving (\ref{eq:vx_SCM_if1}) gives an inversion level $i_{f2}$ that decreases with temperature [Fig.~\ref{fig:3_scm_operation_principle}(b)]. Eq. (\ref{eq:iref}) indicates that, with a proper sizing of the SCM, the PTAT behavior of $I_{SQ}$ and the CTAT one of $i_{f2}$ can cancel each other out, eventually leading to a CWT reference current [Fig.~\ref{fig:3_scm_operation_principle}(c)]. Therefore, the key principle leveraged in the proposed reference is that an SCM biased by a PTAT voltage with a CWT offset generates a temperature-independent reference current. A previous study \cite{Lefebvre_2023} reveals that an $I_{REF}$ TC valley is obtained for a linear relationship between parameters $K_{PTAT}$ and $\alpha$, and that $S_{I_{REF}}$ is improved at higher $V_{off}$ as $M_1$ is biased closer to strong inversion.

\subsection{4T Ultra-Low-Power Voltage Reference}
\label{subsec:2B_4T_ultra_low_power_voltage_reference}
To generate the PTAT voltage with a CWT offset required by the SCM, the proposed reference employs a novel 4T voltage reference architecture, inspired by decades of advances in 2T voltage references \cite{Adriaensen_2002, Seok_2012, CamposDeOliveira_2018, Fassio_2021}, and sharing similarities with 4T structures introduced in \cite{Seok_2012, Huang_2023}. The transistors constituting this 4T voltage reference [Fig.~\ref{fig:2_basic_schematic}(c)] operate in deep subthreshold, in which the drain-to-source current of an nMOS transistor can be described by
\begin{equation}
	I_{DS} = I_{SQ}S \exp\left(\frac{V_{GS}-V_T}{nU_T}\right)\textrm{,}\label{eq:ids_subthreshold}
\end{equation}
for $V_{DS} > 4U_T$, where $I_{SQ} = \mu C_{ox}^{'}(n-1)U_T^2$ is another definition of the specific sheet current, and $V_T$ is the threshold voltage at any $V_{BS}$. In addition, the proposed reference makes extensive use of the body effect, i.e., the change of threshold voltage due to a non-zero $V_{BS}$ voltage, which is captured by
\begin{IEEEeqnarray}{RCL}
	\Delta V_T = V_T - V_{T0} & = &\gamma_b\left(\sqrt{2\phi_{fp}-V_{BS}}-\sqrt{2\phi_{fp}}\right)\textrm{,}\label{eq:body_effect_bulk}\\
	& \approx & -\gamma_b^{*} V_{BS}\IEEEnonumber\textrm{.}
\end{IEEEeqnarray}
in a bulk technology, with $\gamma_b$ the body factor, $\gamma_b^{*}$ its linearization around $V_{BS} = 0$, and $\phi_{fp}$ Fermi's potential, or by
\begin{equation}
	\Delta V_T = V_T - V_{T0} = -\gamma_b^{*} V_{BS}\label{eq:body_effect_fdsoi}
\end{equation}
in an \mbox{FD-SOI} technology, for which $\gamma_b^{*} = C_d/C_{ox}$ is temperature-independent at first order \cite{daSilva_2021} as $C_d$ is the capacitance between the back-gate and channel due to the buried oxide. In addition, it should be noted that in subthreshold, a common expression relating $\gamma_b^{*}$ to the subthreshold slope factor $n$ is given by
\begin{equation}
	\gamma_b^{*} = n-1\textrm{.}\label{eq:gamma_b}
\end{equation}
\indent A simple way to understand the proposed 4T voltage reference [Fig.~\ref{fig:2_basic_schematic}(c)] is to see it as two drain-to-source leakage current sources, corresponding to transistors $M_7$ and $M_9$ connected with a zero $V_{GS}$, biasing $M_6$ to generate a reference voltage $V_X = V_{GS6}$. It should be noted that $I_{DS7} > I_{DS9}$ due to the 2T voltage reference formed by $M_{8-9}$, which leads to $V_{BS7} > 0$ and thus to a forward body biasing (FBB) of $M_7$, ultimately resulting in  a reduction of $V_{T7}$. Applying Kirchhoff's current law at node $V_Y$ gives $I_{DS6} = I_{DS7} + I_{DS9}$. Using the expression of the subthreshold current in (\ref{eq:ids_subthreshold}), and assuming that $M_{6-7-9}$ have the same $I_{SQ}$ and $V_{T0}$, voltage $V_X$ can consequently be expressed as
\begin{equation}
	V_X(T) = n U_T \log\left(\frac{S_9}{S_6} + \frac{S_7}{S_6}\exp\left(\frac{-\Delta V_{T7}}{n U_T}\right)\right)\textrm{,}\label{eq:vx_4T_vref}
\end{equation}
where the decrease in threshold voltage $\Delta V_{T7} < 0$ results in a positive CWT offset $V_{off}$. Moreover, $\Delta V_{T7}$ can be related to $M_7$'s body-to-source voltage through (\ref{eq:body_effect_bulk}) or (\ref{eq:body_effect_fdsoi}), with $V_{BS7}$ found by equating the subthreshold currents $I_{DS8} = I_{DS9}$ in the 2T voltage reference formed by $M_{8-9}$ as in \cite{Seok_2012}, giving
\begin{equation}
	V_{BS7} = \left(\frac{n_8}{n_9}V_{T08} - V_{T09}\right) + n_8 U_T \log\left(\frac{I_{SQ9}}{I_{SQ8}} \frac{S_9}{S_8}\right)\textrm{.}\label{eq:vbs7}
\end{equation}
It should be noted that this FBB technique can be applied in bulk and FD-SOI technologies, with a more limited voltage range in bulk. Different subthreshold factors are used for $M_8$ and $M_9$ as they can be of different $V_T$ types. Next, the nonlinear expression of $V_X$ given by (\ref{eq:vx_4T_vref}) can be expressed as a voltage of the form (\ref{eq:vx_generic}) using a first-order Taylor series approximation of $V_X$ around a temperature $T_0$, provided by
\begin{IEEEeqnarray}{RCL}
	V_X(T) & \approx & V_X(T_0) + \frac{dV_X(T_0)}{dT}\left(T-T_0\right)\textrm{,}\IEEEnonumber\\
	& \approx & V_{X0} + \delta_{V_X}T\textrm{,}
\end{IEEEeqnarray}
with $V_{X0}$ the voltage at zero absolute temperature and $\delta_{V_X}$ the PTAT slope of $V_X$. In practice, we find it more convenient to define quantities which can be extracted from simulation results. In what follows, the CWT offset of $V_X$, denoted as $V_{off}$, is thus defined as the difference between (\ref{eq:vx_4T_vref}) and a purely PTAT reference voltage constituted solely of $M_6$ and $M_9$, at $T_0$ = 25$^\circ$C, and expressed as
\begin{IEEEeqnarray}{RCL}
	V_{off} & \triangleq &V_X(T_0) - n U_{T0} \log\left(\frac{S_9}{S_6}\right)\textrm{,}\IEEEnonumber\\
	& = &n U_{T0} \log\left(1 + \frac{S_7}{S_9}\exp\left(\frac{-\Delta V_{T7}}{n U_{T0}}\right)\right)\textrm{.}
\end{IEEEeqnarray}
It is important to note that $V_{off} \neq V_{X0}$, even though they both represent an offset voltage. Besides, the PTAT slope of $V_X$ is simply estimated as $\Delta V_X/\Delta T \simeq \delta_{V_X}$. Next, Fig.~\ref{fig:5_4T_voltage_reference_generic} illustrates the impact of changes of $S_7/S_6$ and $S_9/S_6$ on voltage $V_X$. First, an increase in $S_7/S_6$ for a fixed value of $S_9/S_6$ [Fig.~\ref{fig:5_4T_voltage_reference_generic}(a)] results in an increased $V_{off}$ ranging from 5 to 55~mV [Fig.~\ref{fig:5_4T_voltage_reference_generic}(b)], while only slightly impacting the PTAT slope [Fig.~\ref{fig:5_4T_voltage_reference_generic}(c)]. Next, an increase in $S_9/S_7$ for a fixed value of $S_7/S_6$ [Fig.~\ref{fig:5_4T_voltage_reference_generic}(d)] strongly impacts the PTAT slope, with a variation from 0.1 to 0.3~mV/$^\circ$C. Changing the ratios $S_7/S_6$ and $S_9/S_6$ therefore allows to tune the CWT offset and PTAT slope of $V_X$ with a relative independence, even though a perfectly independent control cannot be achieved.

\subsection{Proposed Temperature-Independent Current Reference}
\label{subsec:2C_proposed_temperature_independent_current_reference}
\begin{figure}[!t]
	\centering
	\includegraphics[width=.45\textwidth]{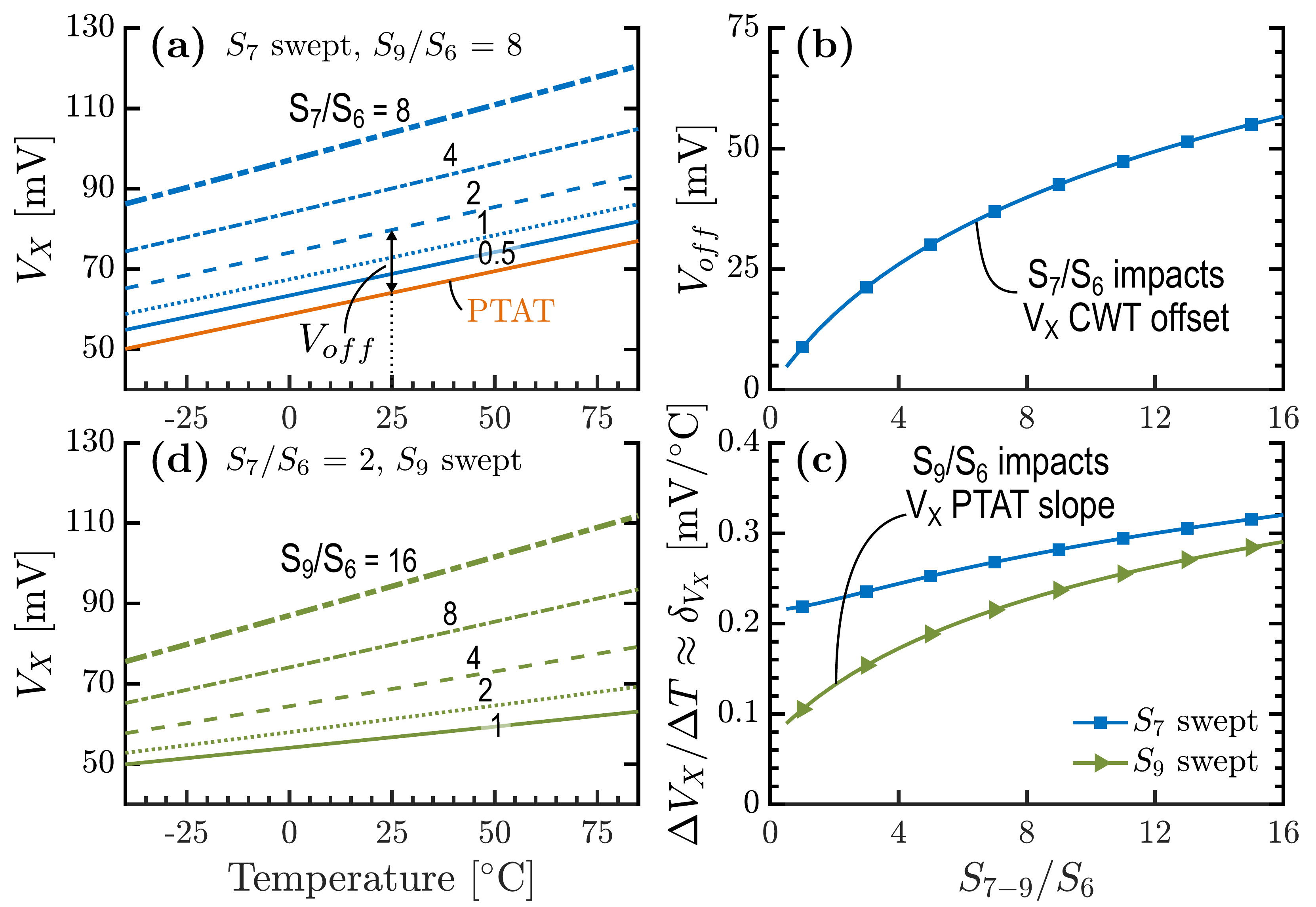}
	\caption{All figures correspond to generic technological parameters $n$~=~1.2, $\gamma_b^{*}$~=~0.15, and $V_{BS7}$~=~0.2~V, and rely on a body effect model corresponding to an \mbox{FD-SOI} technology. Temperature dependence of $V_X$ for (a) a sweep of $S_7$ and a fixed $S_9/S_6$ = 8, and (d) a sweep of $S_9$ and a fixed $S_7/S_6$ = 2. (b) $V_X$ CWT offset for the $S_7$ sweep in (a), and (c) $V_X$ PTAT slope for the sweeps shown in (a) and (d), estimated as the variations of $V_X$ across the temperature range divided by the temperature range.}
	\label{fig:5_4T_voltage_reference_generic}
\end{figure}
The proposed temperature-independent current reference combines two key ideas. Firstly, the idea of biasing an SCM using a PTAT voltage with a CWT offset to generate a CWT current, introduced in \cite{Lefebvre_2023}. This concept shares some similarities with \cite{Osaki_2010}, in which the temperature dependence of the current in a deep-triode transistor is compensated with a temperature-dependent bias voltage. Then, the idea of using a ULP voltage reference whose output is buffered on a $V$-to-$I$ converter, here an SCM, by a single transistor \cite{Wang_2018, Zhuang_2020, Lefebvre_2022}. The main contributions of this work are thus to combine these two concepts and to propose a novel 4T voltage reference architecture to generate the PTAT voltage with a CWT offset required by the SCM.
In addition, the proposed design corrects several limitations of \cite{Lefebvre_2023}, which was using a modified subthreshold $\beta$-multiplier to generate the SCM bias voltage. The four following limitations have been corrected:\\
\indent (i) A $\beta$-multiplier has two stable operating points and requires a startup circuit to ensure that it is biased around the non-zero one. On the contrary, the proposed reference has a single non-zero operating point, eliminating the need for such a circuit. When $I_{REF}$ is equal to zero, $V_X$ and $V_Y$ are $\approx$ 0, but this operating point is unstable as $M_7$ and $M_9$ have a $V_{DS}$ close to $V_{DD}$ and are biasing $M_6$ with a non-zero current leading to $V_{GS6} = V_X > 0$;\\
\indent (ii) The body connection of $M_2$ to $V_X$ in \cite{Lefebvre_2023} degrades $I_{REF}$ TC due to the leakage through the parasitic p-well/n-well diode at node $V_X$. The proposed design solves this issue by connecting the body of $M_2$ to ground, thus deleting this parasitic diode while slightly changing the SCM's equations;\\
\indent (iii) An $I_{REF}$ TC calibration circuit, necessary to maintain an acceptable TC in all process corners, can only be integrated in the previous reference as a binary-weighted current mirror at the cost of significant silicon area to achieve a sufficient resolution, but is rather straightforward to add to the proposed reference with a reasonable area overhead by tuning the effective width of either $M_7$ or $M_9$ through a digital code. A complete explanation of this calibration is provided in Section~\ref{subsec:4A_IREF_TC_calibrationand_overview_of_the_designs}. This feature of the 4T voltage reference is what allows to reduce the measured TC from 565 down to 82~ppm/$^\circ$C for the proposed design in \mbox{22-nm} \mbox{FD-SOI};\\
\indent (iv) The previous reference can only generate a fixed technology-dependent CWT offset related to the body factor and the $\Delta V_T$ between two transistors of different $V_T$ types. There is consequently no degree of freedom to tune $V_{off}$ apart from the transistor type. This issue is crucial as a too-low $V_{off}$ leads to a large sensitivity $S_{I_{REF}}$ and thereby, to a degraded performance in terms of LS and variability of $I_{REF}$ necessitating power and area overheads to mitigate it. Four performance metrics are thus negatively impacted by this lack of tunability. Yet, the proposed circuit allows to adjust $V_{off}$ by simply changing the ratio $S_7/S_6$, as illustrated in Figs.~\ref{fig:5_4T_voltage_reference_generic}(a) and (b), making it possible to optimize $V_{off}$ and to reliably reach acceptable performance in any technology.\\
\indent Solving limitations (iii) and (iv), which are particularly critical regarding the overall performance achieved by the current reference in terms of LS, TC, silicon area, and variability, is only possible because of the proposed 4T voltage reference. This circuit is thus a key enabler of the high level of performance achieved in this work and a significant improvement compared to \cite{Lefebvre_2023}, which cannot be considered as a mere implementation detail.
Nonetheless, the drawback of the proposed current reference is that the power consumption of the 4T voltage reference scales with the subthreshold $I_{DS}$ leakage, and thus increases exponentially with temperature, leading to a larger power consumption than the $\beta$-multiplier-based reference at high temperature.

\section{Design and Sizing Methodology}
\label{sec:3_design_and_sizing_methodology}
\subsection{Overview of the Methodology}
\label{subsec:3A_overview_of_the_methodology}
\begin{figure}[!t]
	\centering
	\includegraphics[width=.45\textwidth]{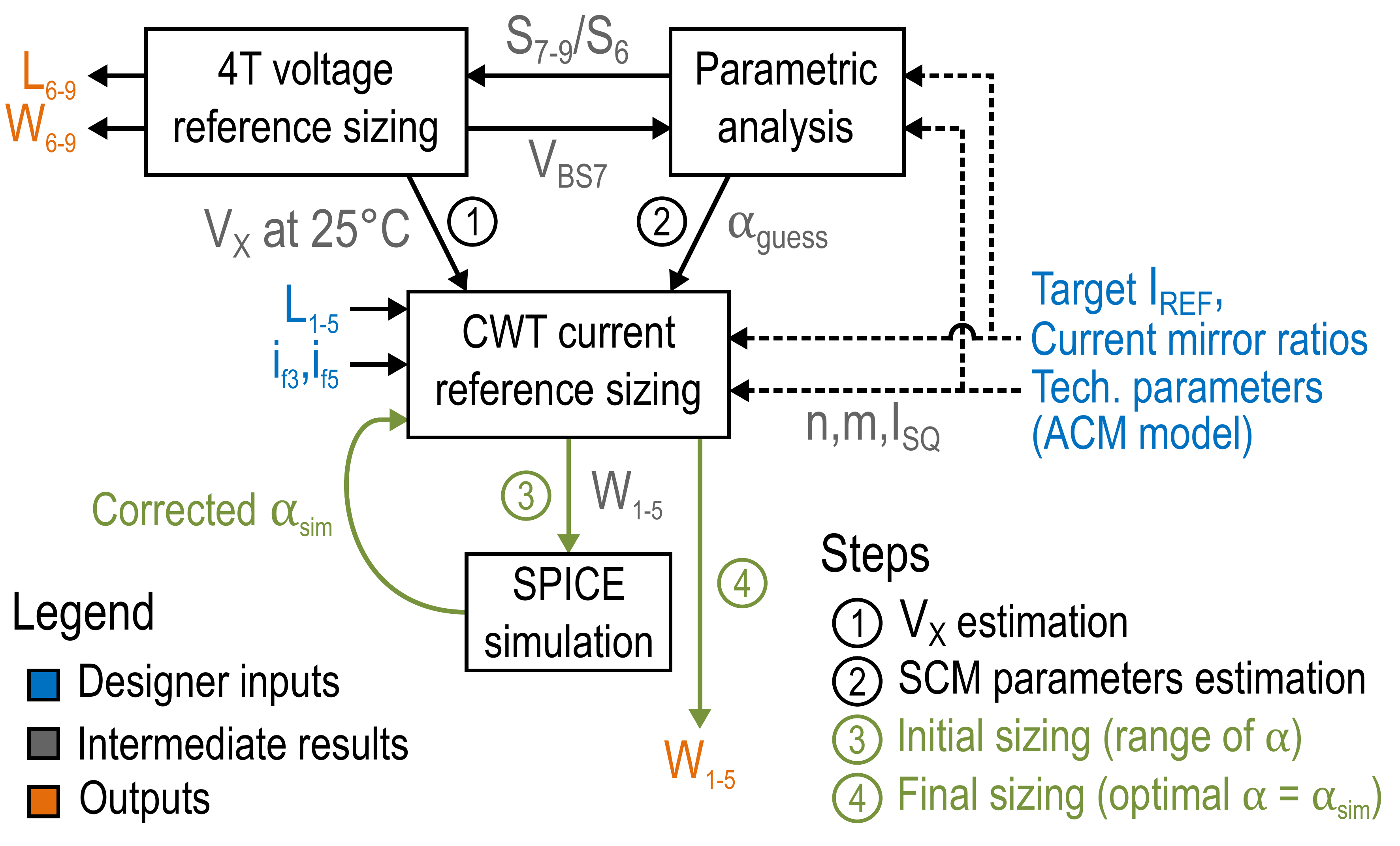}
	\caption{Four-step flowchart of the design and sizing methodology.}
	\label{fig:7_sizing_methodology}
\end{figure}

A flowchart of the sizing methodology is presented in Fig.~\ref{fig:7_sizing_methodology}. It is based on designer inputs such as the target reference current, the ratio of current mirrors, transistor lengths and inversion levels, as well as technological parameters estimated by fitting the ACM model to $g_m/I_D$ curves extracted from SPICE simulations, following the procedure described in \cite{Jespers_2017}. This methodology outputs transistor dimensions and can be divided into four main steps, with \mbox{steps 1)} and 2) being more iterative than sequential.
\par\mbox{Step 1)} sizes the 4T voltage reference and provides an estimate of $V_{BS7}$ after sizing $W_{8-9}$ to make it CWT.
\par\mbox{Step 2)} provides an educated guess for the value of $\alpha$ that minimizes $I_{REF}$ TC, denoted as $\alpha_{\textrm{guess}}$, for a fixed $S_7/S_6$ and a given $S_9/S_6$.
\par\mbox{Step 3)} takes the estimated $V_X$ at 25$^\circ$C from \mbox{step 1)} and $\alpha_{\textrm{guess}}$ from \mbox{step 2)} to compute initial sizings of the current reference for a range of $\alpha$ values and fixed $S_{7-9}/S_6$ ratios in the TT process corner. These sizings, characterized by different $W_{1-2}$ values, are used to run pre-layout SPICE simulations to extract the correct optimal value of $\alpha$, denoted as $\alpha_{\textrm{sim}}$, taking into account the exact transistors' behavior.
\par\mbox{Step 4)} simply consists in running the sizing algorithm with $\alpha = \alpha_{\textrm{sim}}$, thereby generating the final current reference sizing.

\subsection{4T Ultra-Low-Power Voltage Reference}
\label{subsec:3B_4T_ultra-low-power_voltage_reference}
\begin{table*}[!t]
\centering
\caption{Sizing of the proposed nA-range CWT current references.}
\label{table:final_implementation_sizes}
\scalebox{.8}{%
\begin{threeparttable}
\begin{scriptsize}
\begin{tabular}{lcccccccccccc}
	\toprule
	& \multicolumn{6}{c}{UMC \mbox{0.11-$\mu$m} bulk\tnote{$\ast$}} & \multicolumn{6}{c}{GF \mbox{22-nm} FD-SOI}\\
	\cmidrule(lr){2-7} \cmidrule(lr){8-13}
	& & \multicolumn{3}{c}{w/o $I_{REF}$ TC calib.} & \multicolumn{2}{c}{w/ $I_{REF}$ TC calib.} & & \multicolumn{3}{c}{w/o $I_{REF}$ TC calib.} & \multicolumn{2}{c}{w/ $I_{REF}$ TC calib.}\\
	\cmidrule(lr){3-5} \cmidrule(lr){6-7} \cmidrule(lr){9-11} \cmidrule(lr){12-13}
	& Type\tnote{$\star$} & $W$ [$\mu$m] & $L$ [$\mu$m] & $i_f$ & $W$ [$\mu$m] & $L$ [$\mu$m] & Type\tnote{$\star$} & $W$ [$\mu$m] & $L$ [$\mu$m] & $i_f$ & $W$ [$\mu$m] & $L$ [$\mu$m]\\
	\midrule
	$M_1$ & HS & \textbf{0.915}\tnote{$\dagger$} & 40$\times$30\tnote{$\diamond$} & 250.60 & \textbf{0.915}\tnote{$\dagger$} & 40$\times$30\tnote{$\diamond$} & SLVT & \textbf{0.44}\tnote{$\dagger$} & 64$\times$8\tnote{$\diamond$} & 165.82 & \textbf{0.415}\tnote{$\dagger$} & 64$\times$8\tnote{$\diamond$}\\
	$M_2$ & HS & \textbf{0.36}\tnote{$\dagger$} & 40$\times$30\tnote{$\diamond$} & 164.33 & \textbf{0.36}\tnote{$\dagger$} & 40$\times$30\tnote{$\diamond$} & SLVT & \textbf{0.215}\tnote{$\dagger$} & 64$\times$8\tnote{$\diamond$} & 100.50 & \textbf{0.202}\tnote{$\dagger$} & 64$\times$8\tnote{$\diamond$}\\
	$M_3$ & HS & 6$\times$1.25 & 5$\times$10\tnote{$\diamond$} & 1.43 & 6$\times$1.25 & 5$\times$10\tnote{$\diamond$} & LVT & 6$\times$5 & 5$\times$1\tnote{$\diamond$} & 0.02 & 6$\times$5 & 5$\times$1\tnote{$\diamond$}\\
	$M_4$ & HS & 2$\times$1.25 & 5$\times$10\tnote{$\diamond$} & 1.43 & 2$\times$1.25 & 5$\times$10\tnote{$\diamond$} & LVT & 2$\times$5 & 5$\times$1\tnote{$\diamond$} & 0.02 & 2$\times$5 & 5$\times$1\tnote{$\diamond$}\\
	$M_5$ & HS & 1.2 & 10 & 0.10 & 1.2 & 10 & SLVT & 10 & 0.5 & 0.0007 & 10 & 0.5\\
	$M_6$ & HS & 2$\times$2 & 10 & - & 4$\times$2 & 20 & SLVT & 2$\times$5.47 & 1 & - & 2$\times$5.6 & 1\\
	$M_7$ & HS & 8$\times$2 & 10 & - & 16$\times$2 & 20 & SLVT & 4$\times$5.47 & 1 & - & 10$\times$1.12 & 1\\
	$M_{7B}$ & HS & - & - & - & 0.16 & 25 & SLVT & 4$\times$5.47 & 1 & - & 2$\times$1.12 & 1\\
	$M_8$ & LL & 8$\times$2 & 10 & - & 2$\times$2 & 20 & LVT & 16$\times$1.25 & 1 & - & 18$\times$1.28 & 1\\
	$M_9$ & HS & 8$\times$2 & 10 & - & 2$\times$2 & 20 & SLVT & 16$\times$5.47 & 1 & - & 18$\times$5.6 & 1\\
	\midrule
	$M_{SWi}$ & HS & - & - & - & 0.16 & 25 & ULL & - & - & - & 0.16 & 5$\times$8\tnote{$\diamond$}\\
	$M_{7Vi}$ & - & - & - & - & - & - & SLVT & - & - & - & 1 to 16$\times$1.12 & 1\\
	$M_{9Vi}$ & HS & - & - & - & 1 to 16$\times$2 & 20 & - & - & - & - & - & -\\
	\bottomrule
\end{tabular}
\end{scriptsize}
\begin{footnotesize}
\begin{tablenotes}
	\item[$\ast$] Dimensions reported for UMC \mbox{0.11-$\mu$m} bulk are pre-shrink ones, and must be scaled by a factor 0.9$\times$ to obtain silicon dimensions.
	\item[$\star$] In 0.11~$\mu$m, HS refers to high-speed, i.e., LVT, and LL to low-leakage, i.e., HVT. In 22~nm, SLVT refers to super-low-$V_T$, LVT to low-$V_T$, and ULL to ultra-low-leakage, i.e., HVT.
	\item[$\dagger$] Widths fine-tuned based on the outputs of the sizing algorithm to obtain the target $I_{REF}$.
	\item[$\diamond$] $n\times L$ corresponds to a composite transistor, implemented with $n$ transistors of length $L$~$\mu$m connected in series.
\end{tablenotes}
\end{footnotesize}
\end{threeparttable}
}
\end{table*}
The objectives of sizing the voltage reference, corresponding to \mbox{step 1)}, are twofold. First, it aims at selecting a transistor type and length that limit the nonidealities of the voltage reference while minimizing power consumption. Second, it strives to make $V_{BS7}$ CWT by properly sizing $W_{8-9}$.\\
\indent First, the 4T voltage reference can deviate from its ideal behavior at low temperature in the slow nMOS process corners. This behavior is explained by the fact that the subthreshold leakage ceases to decrease exponentially with temperature, either because of gate leakage becoming of the same order of magnitude as the $I_{DS}$ one, or due to gate-induced drain leakage (GIDL), i.e., the increase of $I_{DS}$ leakage at low $V_{GS}$ and large $V_{DS}$ \cite{Lefebvre_2022}.
\begin{figure}[!t]
	\centering
	\includegraphics[width=.48\textwidth]{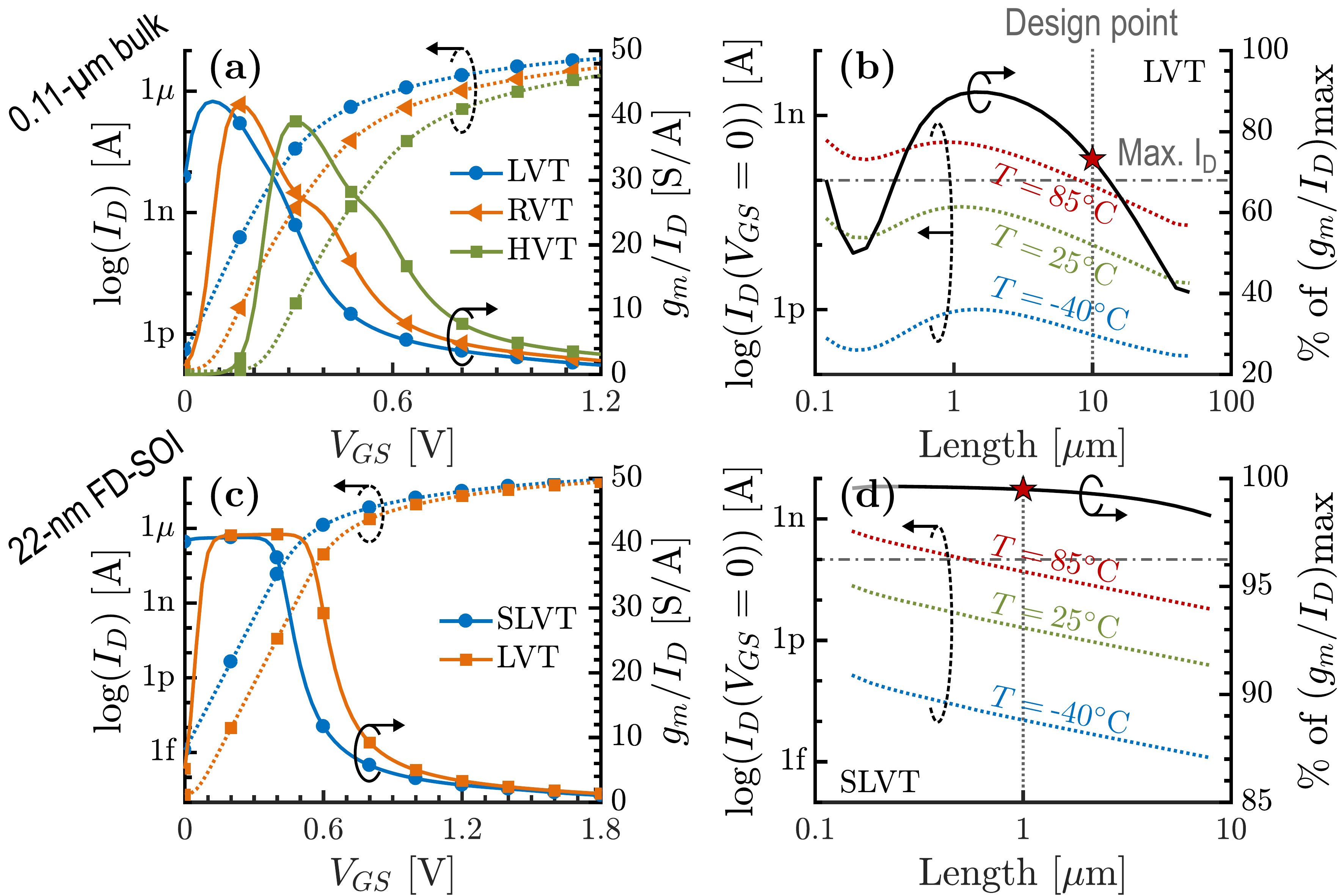}
	\caption{All figures correspond to the SS process corner with a fixed $V_{DS} = V_{GS,\textrm{max}}/2$. $\log(I_D)$, $g_m/I_D$ vs. $V_{GS}$ curves at $T$ = -40$^\circ$C (a) in 0.11~$\mu$m for core LVT, RVT and HVT nMOS with $W$ = 0.5~$\mu$m and $L$ = 10.45~$\mu$m, and (b) in 22~nm for I/O SLVT and LVT nMOS with $W$ = 2~$\mu$m and $L$ = 8~$\mu$m. $\log(I_D)$ at $V_{GS}$ = 0 and at -40, 25 and 85$^\circ$C, and $g_m/I_D$ at $V_{GS}$ = 0 as a percentage of $\left(g_m/I_D\right)_{\textrm{max}}$ (b) in 0.11~$\mu$m for a core LVT nMOS with $W$ = 0.5~$\mu$m and $L$ ranging from 0.12 to 50~$\mu$m, and (d) in 22~nm for an I/O SLVT nMOS with $W$ = 2~$\mu$m and $L$ ranging from 0.15 to 8~$\mu$m.}
	\label{fig:9_sizing_voltage_ref_length}
\end{figure}
\begin{figure}[!t]
	\centering
	\includegraphics[width=.45\textwidth]{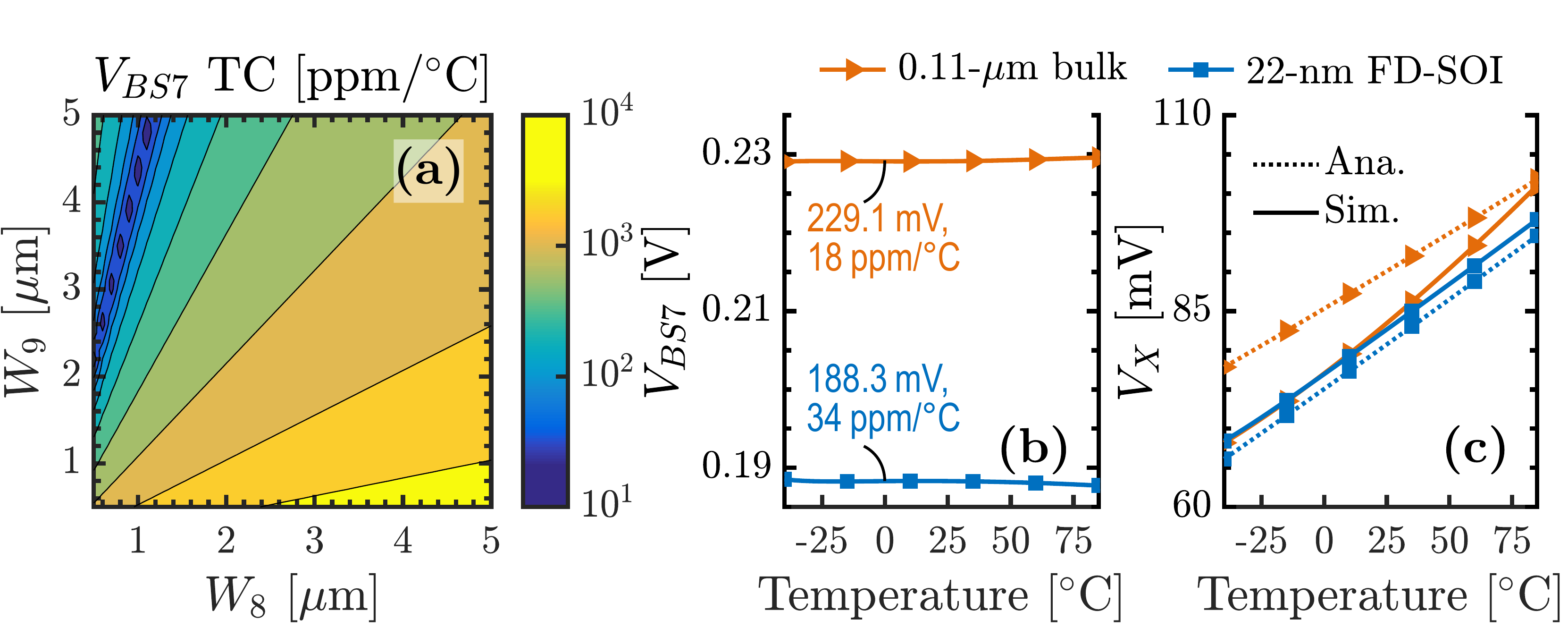}
	\caption{\textbf{$\boldsymbol{V_{BS7}}$ is made CWT by tuning the ratio $\boldsymbol{S_9/S_8}$ in (\ref{eq:vbs7})}. All figures correspond to the TT process corner. (a) $V_{BS7}$ TC from -40 to 85$^\circ$C in 22~nm at 1.8~V, for $W_{8-9}$ swept from 0.5 to 5~$\mu$m and $L_{8-9}$ = 1~$\mu$m. Temperature dependence of (b) $V_{BS7}$ (simulated) and (c) $V_X$ (analytical and simulated), in 0.11~$\mu$m at 1.2~V with $S_7/S_6$ = 4 and $S_9/S_6$ = 4, and in 22~nm at 1.8~V with $S_7/S_6$ = 2 and $S_9/S_6$ = 8.}
	\label{fig:10_sizing_voltage_ref_vx}
\end{figure}
The second explanation is the most likely for the two technologies presented in this work as their gate leakage is negligible. Indeed, \mbox{0.11-$\mu$m} core devices still have a relatively thick oxide, while the devices used in 22~nm are I/O ones with \mbox{high-$\kappa$} gates. Both of these effects are nevertheless captured by a reduced $g_m/I_D$ compared to $\left(g_m/I_D\right)_{\textrm{max}} = 1/(nU_T)$ at $V_{GS}$ = 0. Note that $g_m/I_D$ is simply the slope of the $\log(I_D)$ vs. $V_{GS}$ curve.\\
\indent On the one hand, Figs.~\ref{fig:9_sizing_voltage_ref_length}(a) and (c) depict the $\log(I_D)$ vs. $V_{GS}$ curves in the SS process corner for different transistor types in 0.11~$\mu$m and 22~nm, together with the corresponding $g_m/I_D$ vs. $V_{GS}$ curves. On the other hand, Figs.~\ref{fig:9_sizing_voltage_ref_length}(b) and (d) illustrate the tradeoff between an ideal reference voltage behavior, coinciding with a $g_m/I_D$ at $V_{GS}$ = 0 close to $\left(g_m/I_D\right)_{\textrm{max}}$, and a low power consumption, proportional to $I_D$ at $V_{GS}$ = 0. Here, we set a maximum leakage of 100~pA at 85$^\circ$C for a unitary transistor, but harsher power constraints could be imposed. It should also be observed that the $I_{DS}$ leakage current, and consequently the power consumption, grow exponentially with temperature. In 0.11~$\mu$m, core low-$V_{T}$ (LVT) devices are preferred to regular- or high-$V_{T}$ (RVT or HVT) ones as they are the only transistor type to have an acceptable $g_m/I_D$ drop at $V_{GS}$ = 0 [Fig.~\ref{fig:9_sizing_voltage_ref_length}(a)]. A length of 10~$\mu$m is selected, leading to a $g_m/I_D$ at $V_{GS}$ = 0 and \mbox{-40$^\circ$C} worth 73.2~$\%$ of $\left(g_m/I_D\right)_{\textrm{max}}$, and a \mbox{10.1-pA} $I_D$ at 25$^\circ$C [Fig.~\ref{fig:9_sizing_voltage_ref_length}(b)]. In addition, short-channel effects are visible for small transistor lengths.
\begin{figure}[!t]
	\centering
	\includegraphics[width=.45\textwidth]{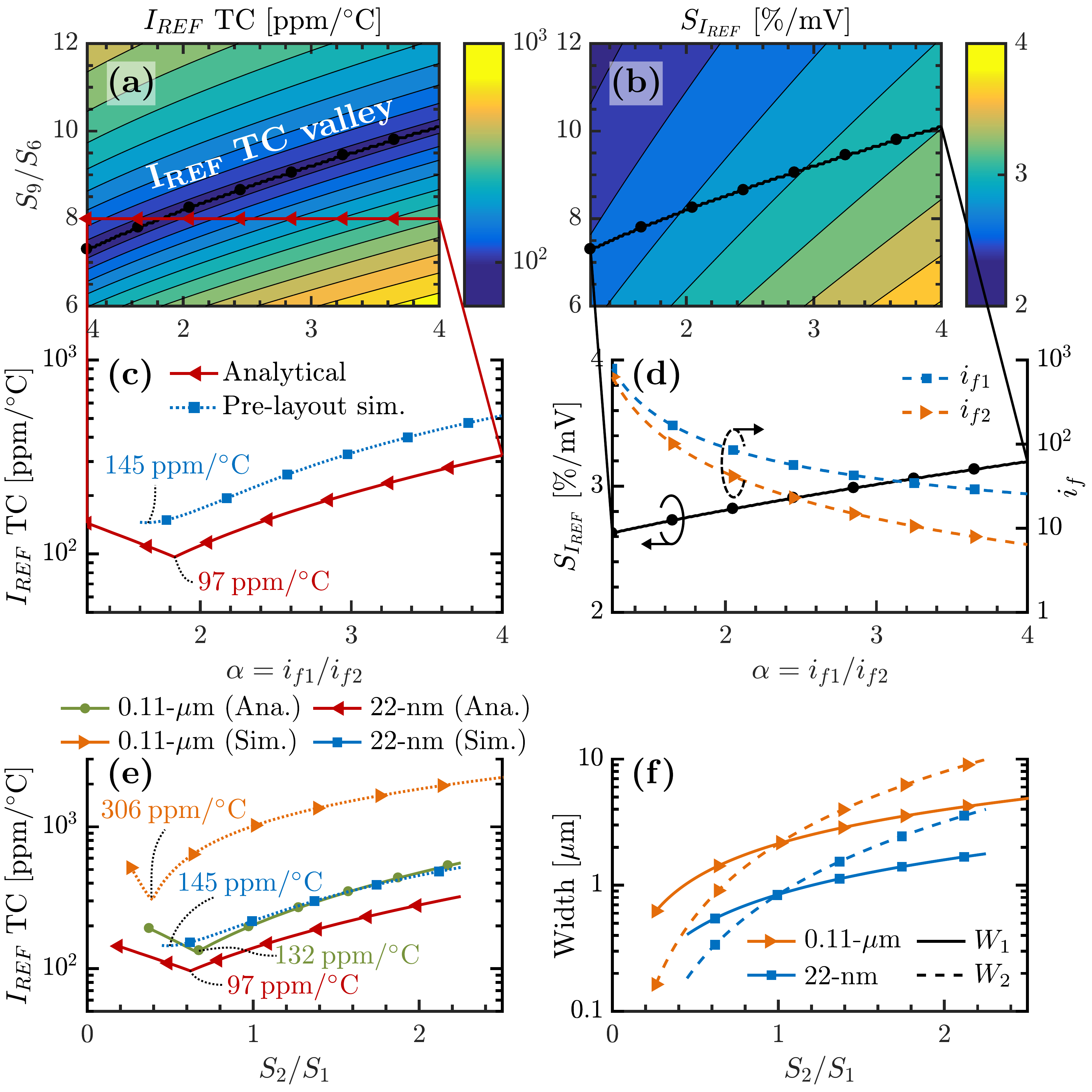}
	\caption{\textbf{$\boldsymbol{I_{REF}}$ is made CWT by properly selecting $\boldsymbol{S_9/S_6}$ and $\boldsymbol{\alpha}$, for a fixed $\boldsymbol{S_7/S_6}$ corresponding to a given CWT offset $\boldsymbol{V_{off}}$.} In 22~nm, (a) $I_{REF}$ TC and (b) $S_{I_{REF}}$ for different values of $(S_9/S_6;\:\alpha)$, with $m$ = 1.63, $n$ = 1.21, and $S_7/S_6$ = 2, corresponding to $V_{off}$ = 17.3~mV. (c) $I_{REF}$ TC, (d) $S_{I_{REF}}$, and $i_{f1-2}$, as a function of $\alpha$ and computed from the analytical model and pre-layout simulations. In 0.11~$\mu$m and 22~nm and as a function of $S_2/S_1$, (e) $I_{REF}$ TC for the analytical model and pre-layout simulations and (f) $W_{1-2}$.}
	\label{fig:11_sizing_current_ref}
\end{figure}
In 22~nm, I/O super-low-$V_T$ (SLVT) devices are picked as they present a perfect behavior at $V_{GS}$ = 0, contrary to LVT ones [Fig.~\ref{fig:9_sizing_voltage_ref_length}(c)]. A length of 1~$\mu$m is chosen, reaching a $g_m/I_D$ at $V_{GS}$ = 0 and \mbox{-40$^\circ$C} of 99.5~$\%$ of $\left(g_m/I_D\right)_{\textrm{max}}$ and a \mbox{2.08-pA} $I_D$ at 25$^\circ$C [Fig.~\ref{fig:9_sizing_voltage_ref_length}(d)].\\
\indent Next, widths $W_{8-9}$ are determined by a simple sweep, as exemplified in Fig.~\ref{fig:10_sizing_voltage_ref_vx}(a) for a \mbox{22-nm} technology. This figure shows that an $S_9/S_8$ ratio of 4.38 leads to a CWT $V_{BS7}$, by having the TC of the second term in (\ref{eq:vbs7}), which is proportional to $U_T$, compensate that of the difference of $V_{T0}$'s between $M_{8-9}$. The resulting pre-layout SPICE simulation results of $V_{BS7}$ and $V_X$ are shown in Figs.~\ref{fig:10_sizing_voltage_ref_vx}(b) and (c), together with analytical expression (\ref{eq:vx_4T_vref}) for $V_X$. In 0.11~$\mu$m, $M_8$ and $M_9$ are respectively implemented with core HVT and LVT nMOS, yielding a \mbox{229-mV} $V_{BS7}$ with an \mbox{18-ppm/$^\circ$C} TC. Regarding $V_X$, the simulation presents a slightly larger PTAT slope than the analytical expression because of the temperature dependence of the body effect in bulk (\ref{eq:body_effect_bulk}) leading to a non-zero $\Delta V_{T7}$ TC. In addition, the body factor $\gamma_b^{*}$ is approximated using (\ref{eq:gamma_b}), thus leading to some discrepancies between the analytical expression and simulation. In 22~nm, $M_8$ and $M_9$ are respectively implemented with I/O LVT and SLVT nMOS, and a \mbox{188-mV} $V_{BS7}$ with a \mbox{34-ppm/$^\circ$C} TC is obtained. $V_X$'s simulated behavior is close to the analytical expression, mostly because $\gamma_b^{*}$ in \mbox{FD-SOI} is temperature-independent at first order \cite{daSilva_2021}, thus making $\Delta V_{T7}$ approximately CWT. The simulation only differs from the analytical expression by a \mbox{1.9-mV} offset.

\subsection{Self-Cascode MOSFET}
\label{subsec:3C_self_cascode_MOSFET}
The parametric analysis of \mbox{step 2)} yields Figs.~\ref{fig:11_sizing_current_ref}(a) and (b). In Fig.~\ref{fig:11_sizing_current_ref}(a), the $I_{REF}$ TC valley corresponds to a quasi-linear relationship between $S_9/S_6$ and $\alpha$. However, Fig.~\ref{fig:11_sizing_current_ref}(b) illustrates that this valley is not an \mbox{iso-$S_{I_{REF}}$} curve as the sizing is performed in the $(S_9/S_6;\:\alpha)$ space instead of the $(K_{PTAT};\:\alpha)$ one. Furthermore, the sizing of the current reference shares similarities with the sizing methodologies proposed in \cite{CamachoGaleano_2005, Lefebvre_2023}. It consists of four main steps:
\begin{enumerate}
	\item Compute voltage $V_X$ at 25$^\circ$C using (\ref{eq:vx_4T_vref}) or use the pre-layout simulation value obtained from \mbox{step 1)};
	\item Solve (\ref{eq:vx_SCM_if1}) for $i_{f2}$, then calculate $i_{f1} = \alpha i_{f2}$, $S_{I_{REF}}$ from (\ref{eq:siref}), $S_2$ from (\ref{eq:id_acm}), and $S_1$ from (\ref{eq:S1_over_S2});
	\item Compute the aspect ratio of transistors $M_{3-4}$ forming the pMOS current mirror using (\ref{eq:id_acm}) while ensuring that $V_{SG4} > 4U_T$;
	\item Compute the aspect ratio of $M_{5}$ using (\ref{eq:id_acm}) while ensuring that $V_Y = V_X + V_{GS5} > 4U_T$.
\end{enumerate}
\begin{figure}[!t]
	\centering
	\includegraphics[width=.5\textwidth]{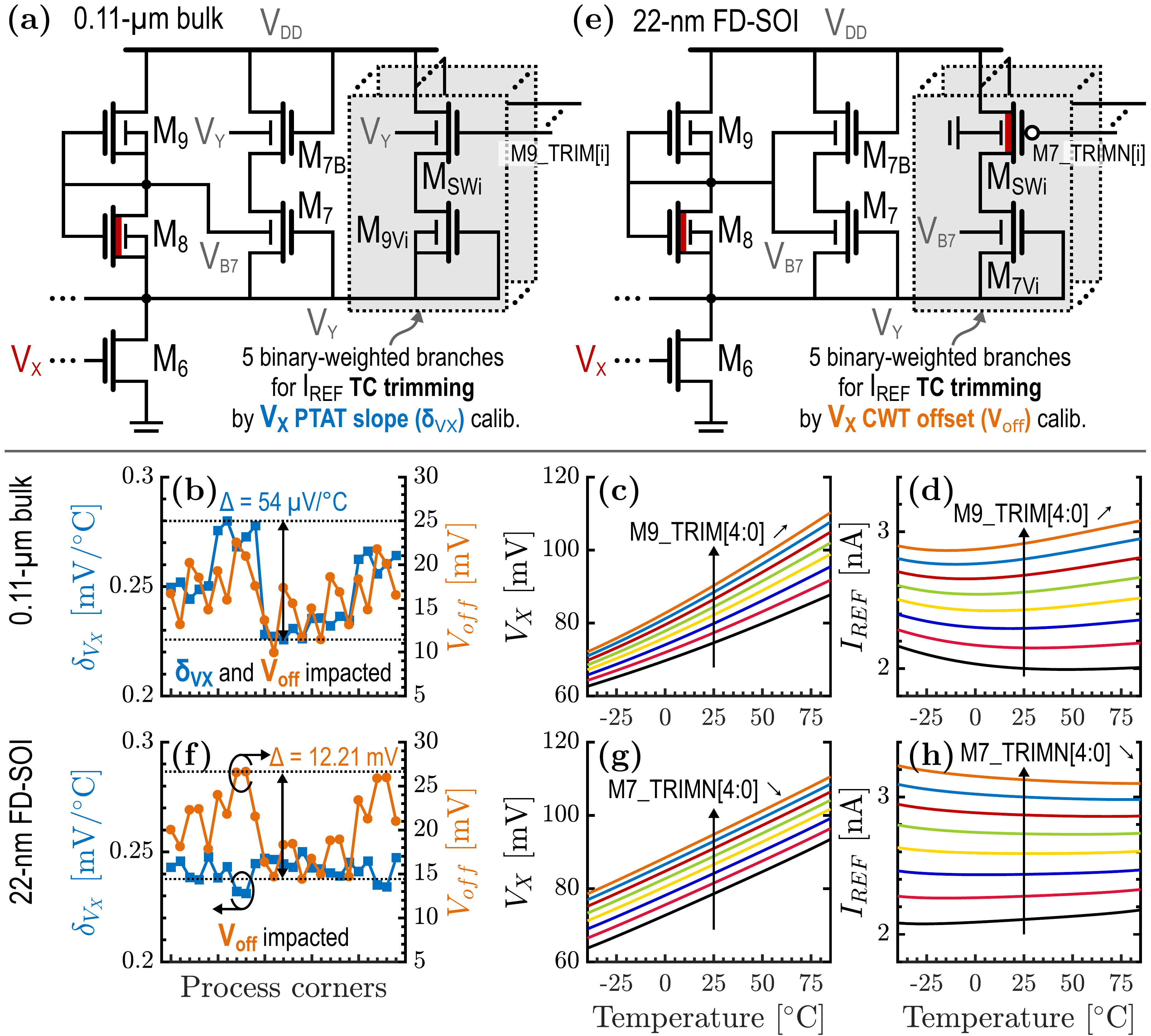}
	\caption{\textbf{$\boldsymbol{I_{REF}}$'s temperature independence can be maintained across process corners by tuning $\boldsymbol{V_X}$ PTAT slope ($\boldsymbol{\delta_{V_X}}$) in \mbox{0.11-$\mu$m} bulk or $\boldsymbol{V_X}$ CWT offset ($\boldsymbol{V_{off}}$) in \mbox{22-nm} FD-SOI}. Schematic of the 4T voltage reference with $I_{REF}$ TC calibration circuit in (a) 0.11~$\mu$m and (b) 22~nm. $\delta_{V_X}$ and $V_{off}$ in process corners (considering skewed process corners of the nMOS of different $V_T$ types used for $M_{8-9}$) in (b) 0.11~$\mu$m and (f) 22~nm. Temperature dependence of $V_X$ and $I_{REF}$ for different \mbox{5-bit} calibration codes in (c)(d) 0.11~$\mu$m and (g)(h) 22~nm, in the TT process corner.}
	\label{fig:12_final_schematic}
\end{figure}
\begin{table}[!t]
\centering
\caption{Summary of the simulated and measured performance of the proposed nA-range CWT references.}
\label{table:summary_performance}
\scalebox{.8}{%
\begin{scriptsize}
\begin{tabular}{lcccccccc}
	\toprule
	& \multicolumn{4}{c}{UMC \mbox{0.11-$\mu$m} bulk} & \multicolumn{4}{c}{GF \mbox{22-nm} FD-SOI}\\
	\cmidrule(lr){2-5} \cmidrule(lr){6-9}
	& \multicolumn{2}{c}{w/o TC calib.} & \multicolumn{2}{c}{w/ TC calib.} & \multicolumn{2}{c}{w/o TC calib.} & \multicolumn{2}{c}{w/ TC calib.}\\
	& Sim. & Meas. & Sim. & Meas. & Sim. & Meas. & Sim. & Meas.\\
	\midrule
	$I_{REF}$ [nA] & 2.49 & 2.74 & 2.40 & 2.30 & 2.48 & 2.32 & 2.50 & 2.54\\
	Power [nW] & 17.03 & 19.37 & 16.65 & 16.82 & 15.43 & 14.44 & 15.49 & 16.30\\
	& $@$1.2V & $@$1.2V & $@$1.2V & $@$1.2V & $@$1.5V & $@$1.5V & $@$1.5V & $@$1.5V\\
	Area [mm$^2$] & \multicolumn{2}{c}{0.00657} & \multicolumn{2}{c}{0.01061} & \multicolumn{2}{c}{0.00222} & \multicolumn{2}{c}{0.00255}\\
	\midrule
	Supply range [V] & \multicolumn{2}{c}{0.8 -- 1.2} & \multicolumn{2}{c}{0.8 -- 1.2} & \multicolumn{2}{c}{1 -- 1.8} & \multicolumn{2}{c}{1 -- 1.8}\\
	LS [$\%$/V] & 2.94 & 2.60 & 2.07 & 2.23 & 0.21 & 1.47 & 0.26 & 1.53\\
	\midrule
	Temp. range [$^\circ$C] & \multicolumn{2}{c}{-40 -- 85} & \multicolumn{2}{c}{-40 -- 85} & \multicolumn{2}{c}{-40 -- 85} & \multicolumn{2}{c}{-40 -- 85}\\
	TC [ppm/$^\circ$C] & 330.7 & 529.5 & 290.6 & 176.0 & 137.5 & 360.6 & 101.5 & 81.5\\
	\midrule
	$I_{REF}$ var. & \multirow{2}{*}{6.00} & \multirow{4}{*}{2.11} & \multirow{2}{*}{7.53} & \multirow{4}{*}{3.47} & \multirow{2}{*}{3.16} & \multirow{4}{*}{2.48} & \multirow{2}{*}{4.18} & \multirow{4}{*}{2.55}\\
	(process) [$\%$] & & & & & & & & \\
	$I_{REF}$ var. & \multirow{2}{*}{1.32} & & \multirow{2}{*}{1.74} & & \multirow{2}{*}{2.68} & & \multirow{2}{*}{2.41} & \\
	(mismatch) [$\%$] & & & & & & & & \\
	\midrule
	$t_{start}$ [ms] & 15.25 & 17.53 & 16.21 & 18.22 & 3.29 & 11.56 & 3.26 & 11.20\\
	\bottomrule
\end{tabular}
\end{scriptsize}
}
\end{table}
\indent Let us now have a closer look at the results of the sizing algorithm shown in Figs.~\ref{fig:11_sizing_current_ref}(c) to (f). First, Fig.~\ref{fig:11_sizing_current_ref}(d) illustrates that there is an interest in choosing small values of $\alpha$ as they result in a lower $S_{I_{REF}}$. This trend stems from the fact that $M_{1-2}$ are biased in moderate ($i_f \in [1;\:100]$) or strong ($i_f > 100$) inversion, at the cost of a larger minimum supply voltage. They will remain close to this inversion level in process corners thanks to the SCM structure, and in temperature corners thanks to the relatively limited variations of $i_{f}$ with temperature [Fig.~\ref{fig:3_scm_operation_principle}(b)]. A design point corresponding to $(S_9/S_6;\:\alpha_\textrm{guess})$ = (8;$\:$1.825) is selected based on Figs.~\ref{fig:11_sizing_current_ref}(a) and (c), yielding an $I_{REF}$ TC = 96.7~ppm/$^\circ$C and $S_{I_{REF}}$ = 2.77~$\%/V$. Pre-layout SPICE simulations in \mbox{steps 3)} and 4) reveal that a \mbox{145-ppm/$^\circ$C} $I_{REF}$ TC is eventually reached for $\alpha_{\textrm{sim}}$ = 1.65, also corresponding to $S_{I_{REF}}$ = 2.73~$\%/V$. Fig.~\ref{fig:11_sizing_current_ref}(e) depicts the same kind of results as Fig.~\ref{fig:11_sizing_current_ref}(c), but as a function of $S_2/S_1$, used as a proxy for $\alpha$ and calculated from (\ref{eq:S1_over_S2}). In 0.11~$\mu$m, the difference between the \mbox{131.5-ppm/$^\circ$C} analytical optimum located at $S_2/S_1$ = 0.66 and the \mbox{306-ppm/$^\circ$C} pre-layout one obtained for $S_2/S_1$ = 0.39 originates from the temperature dependence of the body effect, which was not accounted for in the analytical model, as detailed in Section~\ref{subsec:3B_4T_ultra-low-power_voltage_reference}. In 22~nm, the \mbox{96.7-ppm/$^\circ$C} analytical optimum located at $S_2/S_1$ = 0.62 and the \mbox{145-ppm/$^\circ$C} pre-layout one obtained for $S_2/S_1$ = 0.49 are closer to each other but still differ due to inaccuracies of the ACM model, and more specifically, of the fitting of the temperature exponent of carrier mobility $m$. Finally, Fig.~\ref{fig:11_sizing_current_ref}(f) demonstrates that $W_{1-2}$ decrease with $\alpha$ as $M_{1-2}$ are biased in stronger inversion while keeping $L_{1-2}$ and $I_{D1-2}$ fixed. In addition, the transistor widths are larger in 0.11~$\mu$m because $M_{1-2}$ are $\approx$ 2$\times$ longer in this technology ($L$ = 40$\times$30~$\mu$m = 1.2~mm) than in 22~nm ($L$ = 64$\times$8~$\mu$m = 512 $\mu$m).

\section{Simulation and Measurement Results}
In this section, we present the $I_{REF}$ TC calibration circuit implemented in the proposed reference, and provide an overview of the implemented current references and the measurement setup. Then, we discuss into details the post-layout simulations and measurements of the designs in \mbox{0.11-$\mu$m} bulk and \mbox{22-nm} \mbox{FD-SOI}, whose summary is provided in Table~\ref{table:summary_performance}.

\label{sec:4_simulation_and_measurement_results}
\subsection{$I_{REF}$ TC Calibration and Overview of the Designs}
\label{subsec:4A_IREF_TC_calibrationand_overview_of_the_designs}
\begin{figure}[!t]
	\centering
	\includegraphics[width=.469\textwidth]{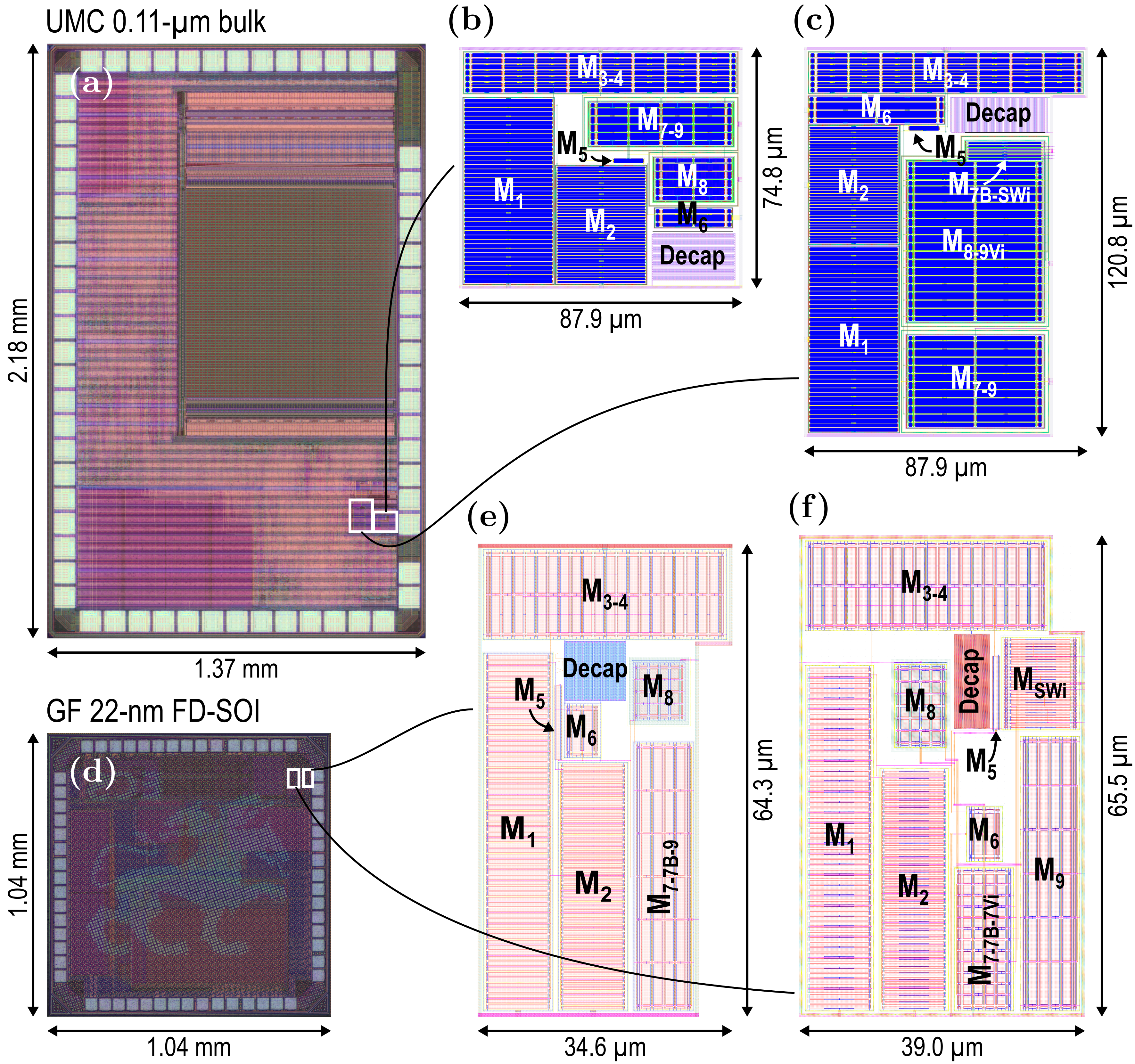}
	\caption{Chip microphotographs with overlaid layout in (a) UMC \mbox{0.11-$\mu$m} bulk and (d) GF \mbox{22-nm} \mbox{FD-SOI}. Layouts of the proposed nA-range CWT current references (b)(e) without and (c)(f) with $I_{REF}$ TC calibration.}
	\label{fig:13_microphotograph_layout}
\end{figure}
Fig.~\ref{fig:12_final_schematic} shows the $I_{REF}$ TC calibration mechanism implemented in the proposed current references. First, in \mbox{0.11-$\mu$m} bulk, process variations (including skewed process corners of $M_{8-9}$) impact $V_X$ PTAT slope ($\delta_{V_X}$) and CWT offset ($V_{off}$) [Fig.~\ref{fig:12_final_schematic}(b)], with total variations of 54~$\mu$V/$^\circ$C and 12.57~mV, respectively. When both quantities are impacted, we observe that calibrating $\delta_{V_X}$ is more effective than calibrating $V_{off}$ in terms of number of components. Tuning the number of transistors placed in parallel to implement $M_9$ using a calibration code [Fig.~\ref{fig:12_final_schematic}(a)] modifies $\delta_{V_X}$ by changing the ratio $S_9/S_6$ in (\ref{eq:vx_4T_vref}) [Fig.~\ref{fig:12_final_schematic}(c)], thus allowing to adjust the $I_{REF}$ TC [Fig.~\ref{fig:12_final_schematic}(d)]. Then, in \mbox{22-nm} \mbox{FD-SOI}, process variations mostly impact $V_X$ CWT offset ($V_{off}$) with a total variation of 12.21~mV, and have a lesser influence on $\delta_{V_X}$ than in 0.11~$\mu$m, as it only changes by 19~$\mu$V/$^\circ$C [Fig.~\ref{fig:12_final_schematic}(f)]. It is then more effective to calibrate $V_{off}$.
\begin{figure}[!t]
	\centering
	\includegraphics[width=.48\textwidth]{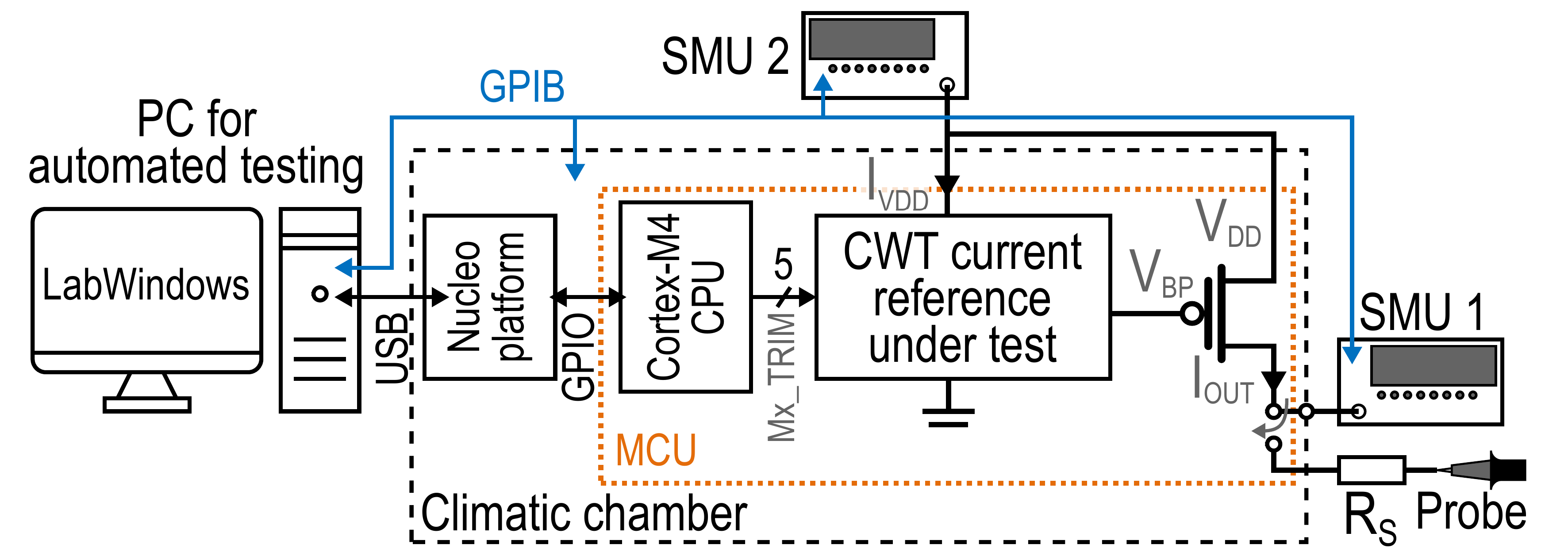}
	\caption{Conceptual measurement testbench for startup time, supply voltage and temperature dependence characterization.}
	\label{fig:14_meas_setup}
\end{figure}
\begin{figure}[!t]
	\centering
	\includegraphics[width=.45\textwidth]{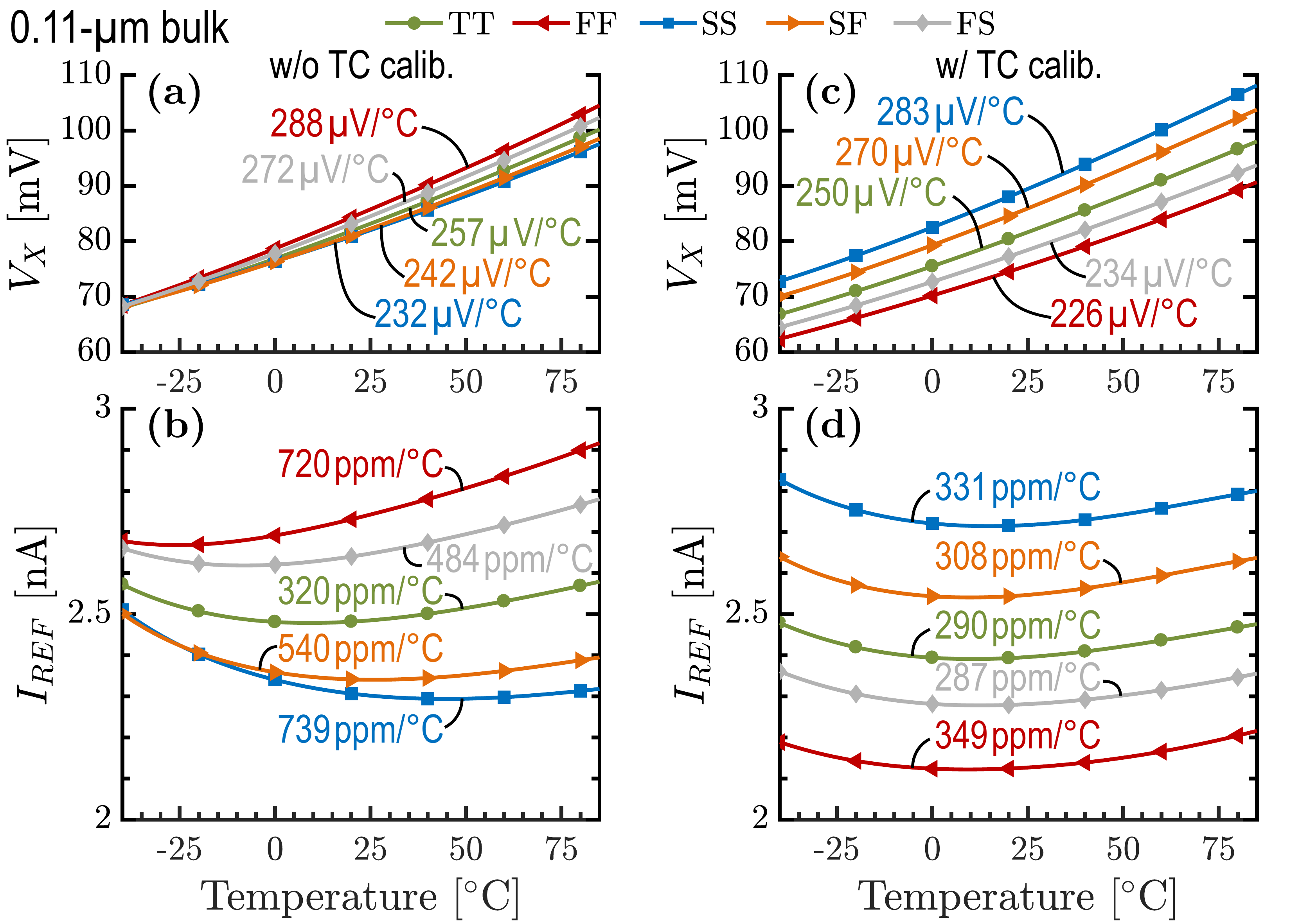}
	\caption{In UMC \mbox{0.11-$\mu$m} bulk, post-layout simulation of the temperature dependence of $V_X$ and $I_{REF}$, in all process corners and at 1.2~V, without [(a) and (b)] and with $I_{REF}$ TC calibration [(c) and (d)].}
	\label{fig:15_sim_iref_vs_T_0p11um}
\end{figure}
\begin{figure}[!t]
	\centering
	\includegraphics[width=.45\textwidth]{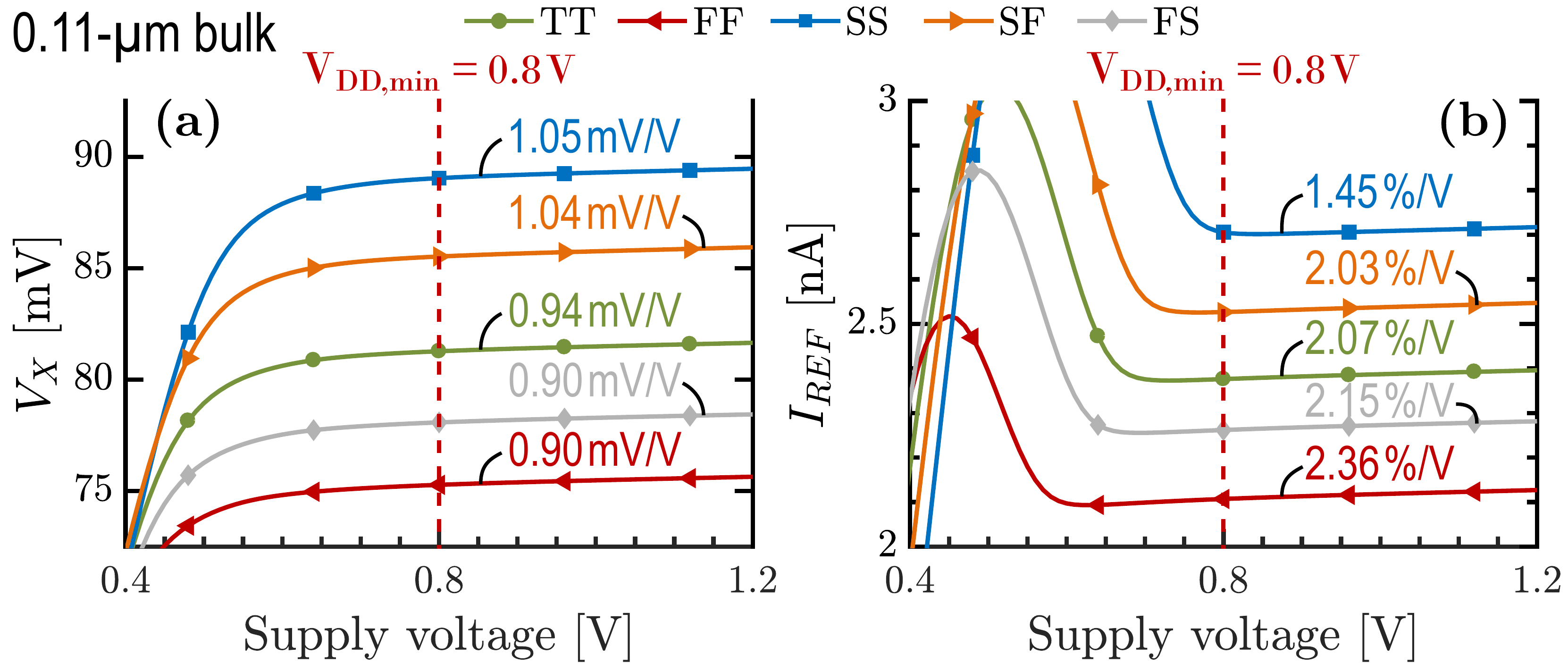}
	\caption{In UMC \mbox{0.11-$\mu$m} bulk, post-layout simulation of the supply voltage dependence of (a) $V_X$ and (b) $I_{REF}$ in all process corners and at 25$^\circ$C, with $I_{REF}$ TC calibration.}
	\label{fig:16_sim_iref_vs_vdd_0p11um}
\end{figure}
Tuning the number of transistors placed in parallel to implement $M_7$ using a calibration code [Fig.~\ref{fig:12_final_schematic}(e)] impacts $V_{off}$ by changing the ratio $S_7/S_6$ in (\ref{eq:vx_4T_vref}) [Fig.~\ref{fig:12_final_schematic}(g)], thus providing another way to calibrate the $I_{REF}$ TC [Fig.~\ref{fig:12_final_schematic}(h)].
Besides, for both technologies, the TC calibration range can be extended to provide a design margin against inaccurate device models. In addition, the impact of junction leakage, gate leakage, or parasitic diode leakage on the 4T voltage reference remains below 1~$\%$ of the total current in all PVT corners, preventing nonidealities and avoiding the need for any circuit adjustment.
Two current references, respectively without and with $I_{REF}$ TC calibration, have been fabricated in each of the UMC \mbox{0.11-$\mu$m} bulk and GF \mbox{22-nm} \mbox{FD-SOI} technology nodes. Chip microphotographs, layouts and final dimensions of these four current references can be found in Fig.~\ref{fig:13_microphotograph_layout} and Table~\ref{table:final_implementation_sizes}, respectively. Based on these references, the area overhead of the the TC calibration circuit is estimated to be 61.5~$\%$ compared to the design without calibration in 0.11~$\mu$m, and 14.9~$\%$ in 22~nm. This gap in area overhead originates from the much-longer transistors used in the TC calibration circuit in 0.11~$\mu$m. Only the power consumption of the 4T voltage reference is affected by the calibration, and the power overhead is thus dependent on the calibration code.

\subsection{Measurement Testbench}
\label{subsec:4B_measurement_testbench}
\begin{figure}[!t]
	\centering
	\includegraphics[width=.45\textwidth]{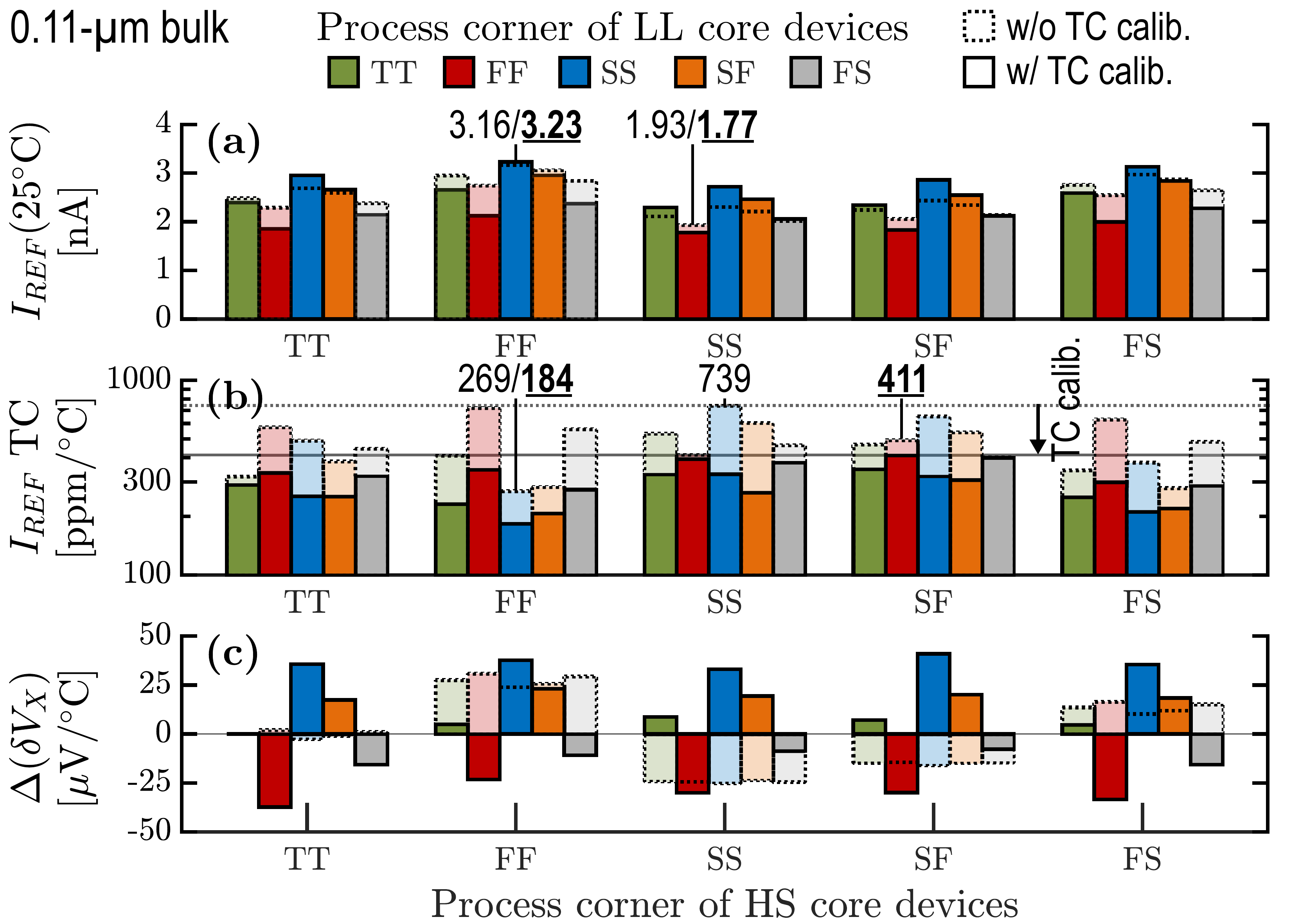}
	\caption{In UMC \mbox{0.11-$\mu$m} bulk and at 1.2~V, post-layout simulation of (a) $I_{REF}$ at 25$^\circ$C, (b) $I_{REF}$ TC from -40 to 85$^\circ$C, and (c) the change in PTAT slope of $V_X$ with respect to its nominal value, without and with TC calibration. Skewed process corners of LL and HS core devices are considered.}
	\label{fig:17_sim_iref_vs_T_skewed_process_0p11um}
\end{figure}
\begin{figure}[!t]
	\centering
	\includegraphics[width=.5\textwidth]{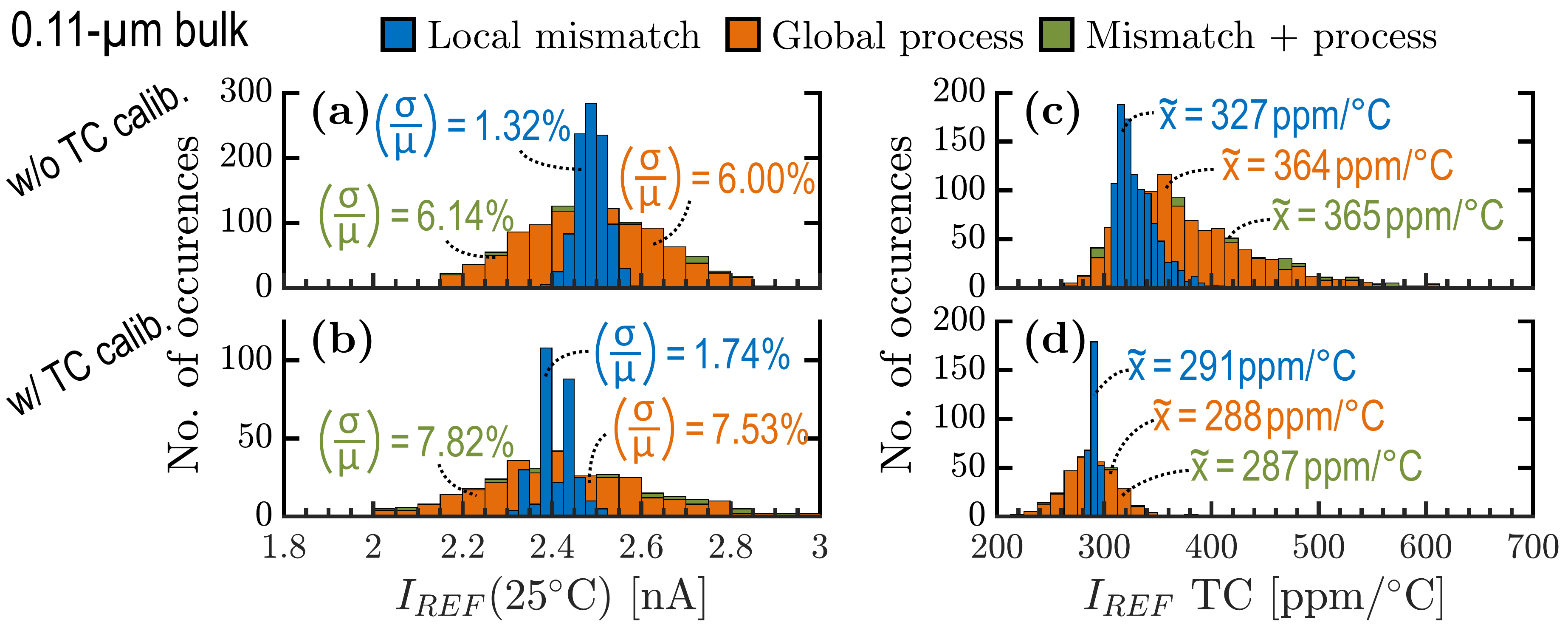}
	\caption{In UMC \mbox{0.11-$\mu$m} bulk, for 10$^3$ / 3$\times$10$^2$ post-layout MC simulations in TT at 1.2~V, histograms of $I_{REF}$ at 25$^\circ$C [(a) and (b)] and of $I_{REF}$ TC from -40 to 85$^\circ$C [(c) and (d)], without [(a) and (c)] and with $I_{REF}$ TC calibration [(b) and (d)]. $\tilde{x}$ denotes the median of a statistical distribution.}
	\label{fig:18_sim_iref_vs_T_mc_0p11um}
\end{figure}
The measurement setup is schematized in Fig.~\ref{fig:14_meas_setup}. The proposed current references are integrated in microcontroller units (MCUs), called (i) MANTIS in \mbox{0.11-$\mu$m} bulk, intended for near-sensor image processing applications, and (ii) CERBERUS in \mbox{22-nm} \mbox{FD-SOI}, intended for edge machine-learning applications and low-power wide-area network communication. Eleven dies have been measured in each technology node. Furthermore, the testing is controlled by a host PC controlling an Espec SH-261 climatic chamber for the temperature sweep, and two Keithley K2450 source measure units (SMUs) for the supply voltage sweep. Note that the stabilization time at each temperature step is sufficiently long to ensure that the die reaches the same temperature as the chamber. Besides, \mbox{SMU 1} measures the output current $I_{OUT}$ with $V_{SD} = V_{DD,\textrm{max}}/2$, respectively equal to 4$\times$ and 8$\times I_{REF}$ in 0.11~$\mu$m and 22~nm, and \mbox{SMU 2}, the sum of $I_{OUT}$ and the supply current $I_{VDD}$. The pMOS current mirror generating $I_{OUT}$ employs a common-centroid layout to accurately replicate $I_{REF}$, and only marginally augments $I_{REF}$ $(\sigma/\mu)$ by 0.03 and 0.06~$\%$ in 0.11~$\mu$m and 22~nm, respectively. In addition, the host PC controls the TC calibration code through a Nucleo platform interacting with the on-chip Cortex-M4 central processing unit (CPU). At last, startup measurements are performed by replacing \mbox{SMU 1} with an 82-\:/$\:$47-M$\Omega$ resistor $R_S$ in series with the 1-M$\Omega$ input resistance of the oscilloscope probe.
\begin{figure}[!t]
	\centering
	\includegraphics[width=.5\textwidth]{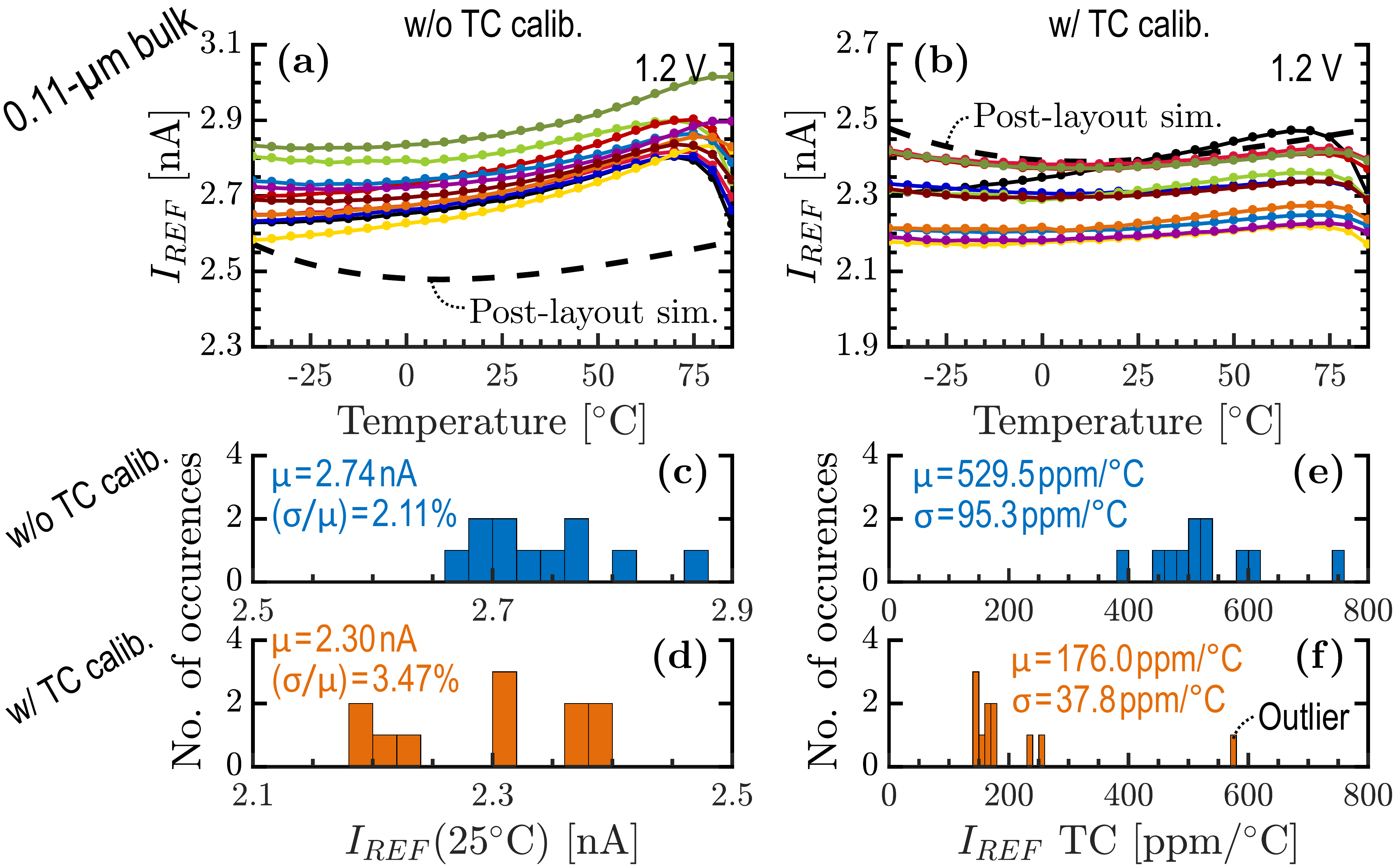}
	\caption{In UMC \mbox{0.11-$\mu$m} bulk and at 1.2~V, measured temperature dependence of $I_{REF}$ (a) without and (b) with $I_{REF}$ TC calibration. Measured histograms of $I_{REF}$ at 25$^\circ$C [(c) and (d)] and of $I_{REF}$ TC from -40 to 85$^\circ$C [(e) and (f)], without and with $I_{REF}$ TC calibration.}
	\label{fig:19_meas_iref_vs_T_0p11um}
\end{figure}

\subsection{Designs in 0.11-$\mu$m Bulk CMOS Technology}
\label{subsec:4C_designs_in_0.11-um_bulk_CMOS_technology}
First, we discuss the post-layout simulation results of the references. Fig.~\ref{fig:15_sim_iref_vs_T_0p11um}(a) reveals that, without TC calibration, $\delta_{V_X}$ spans from 232 to 288~$\mu$V/$^\circ$C due to process variations of the subthreshold slope factor $n$. This results in a \mbox{739-ppm/$^\circ$C} CTAT to \mbox{720-ppm/$^\circ$C} PTAT $I_{REF}$ TC in Fig.~\ref{fig:15_sim_iref_vs_T_0p11um}(b). With TC calibration, $\delta_{V_X}$ is adapted to each process corner [Fig.~\ref{fig:15_sim_iref_vs_T_0p11um}(c)], leading to an $I_{REF}$ TC between 287 and 349~ppm/$^\circ$C with a residual second order temperature dependence [Fig.~\ref{fig:15_sim_iref_vs_T_0p11um}(d)]. Then, Fig.~\ref{fig:16_sim_iref_vs_vdd_0p11um} presents the supply voltage dependence of the reference with TC calibration, whose TC has been calibrated in each process corner. Fig.~\ref{fig:16_sim_iref_vs_vdd_0p11um}(a) shows that $V_X$ LS lies between 0.9 and 1.05~mV/V from 0.8 to 1.2~V.
The analytical expression of the minimum supply voltage is given by
\begin{equation}
	V_{DD,\textrm{min}} = 4U_T + \max (V_G,\:V_X+V_{SG4},\:V_X+V_{GS5}+V_{GS8})\textrm{,} \label{eq:vddmin}
\end{equation}
and $V_{DD,\textrm{min}}$ = 0.8~V is here limited by voltage $V_G$ in the SCM. This translates into an LS of $I_{REF}$ between 1.45 and 2.36~$\%$/V in Fig.~\ref{fig:16_sim_iref_vs_vdd_0p11um}(b). Next, we consider skewed process corners of high-speed (HS), i.e., LVT, and low-leakage (LL), i.e., HVT, core devices in Fig.~\ref{fig:17_sim_iref_vs_T_skewed_process_0p11um}. We thus consider independent process variations for transistors of different $V_T$ types. The TC calibration moderately raises $I_{REF}$ process variations from \mbox{+26.4}~$\%\:$/$\:$\mbox{-22.8}~$\%$ to \mbox{+29.2}~$\%\:$/$\:$\mbox{-29.2}~$\%$ in the (FF, SS) and (SS, FF) corners, respectively. This behavior stems from tuning $\delta_{V_X}$, which affects the value of $V_X$ itself [Fig.~\ref{fig:17_sim_iref_vs_T_skewed_process_0p11um}(a)]. Meanwhile, the calibration shifts the TC from 289 to 739~ppm/$^\circ$C, down to 184 to 411~ppm/$^\circ$C [Fig.~\ref{fig:17_sim_iref_vs_T_skewed_process_0p11um}(b)]. The calibration mechanism tends to harmonize $\delta_{V_X}$ for a given LL process corner, while $\delta_{V_X}$ was initially quite uniform in a given HS process corner [Fig.~\ref{fig:17_sim_iref_vs_T_skewed_process_0p11um}(c)]. Finally, Fig.~\ref{fig:18_sim_iref_vs_T_mc_0p11um}(a) emphasizes that, without TC calibration, $I_{REF}$ varies more significantly due to global process variations, with a \mbox{6-$\%$} $(\sigma/\mu)$, than from local mismatch, with a \mbox{1.32-$\%$} $(\sigma/\mu)$. The same conclusion can be drawn with TC calibration [Fig.~\ref{fig:18_sim_iref_vs_T_mc_0p11um}(b)], with slightly larger $(\sigma/\mu)$'s of 7.53~$\%$ and 1.74~$\%$ for process and mismatch, due to a different sizing of the 4T voltage reference for the TC calibration circuit, coupled with the tendency of the calibration to accentuate $I_{REF}$ variations. Regarding $I_{REF}$ TC, calibration improves the median value $\tilde{x}$ by 40 to 80~ppm/$^\circ$C, but most notably shortens the tail of the distribution, with a 99th percentile for combined process and mismatch reduced from 576 to 364~ppm/$^\circ$C [Figs.~\ref{fig:18_sim_iref_vs_T_mc_0p11um}(c) and (d)].\\
\indent Then, we present the measurement results of the 11 dies. The temperature dependence without and with TC calibration is shown in Figs.~\ref{fig:19_meas_iref_vs_T_0p11um}(a) and (b).
\begin{figure}[!t]
	\centering
	\includegraphics[width=.5\textwidth]{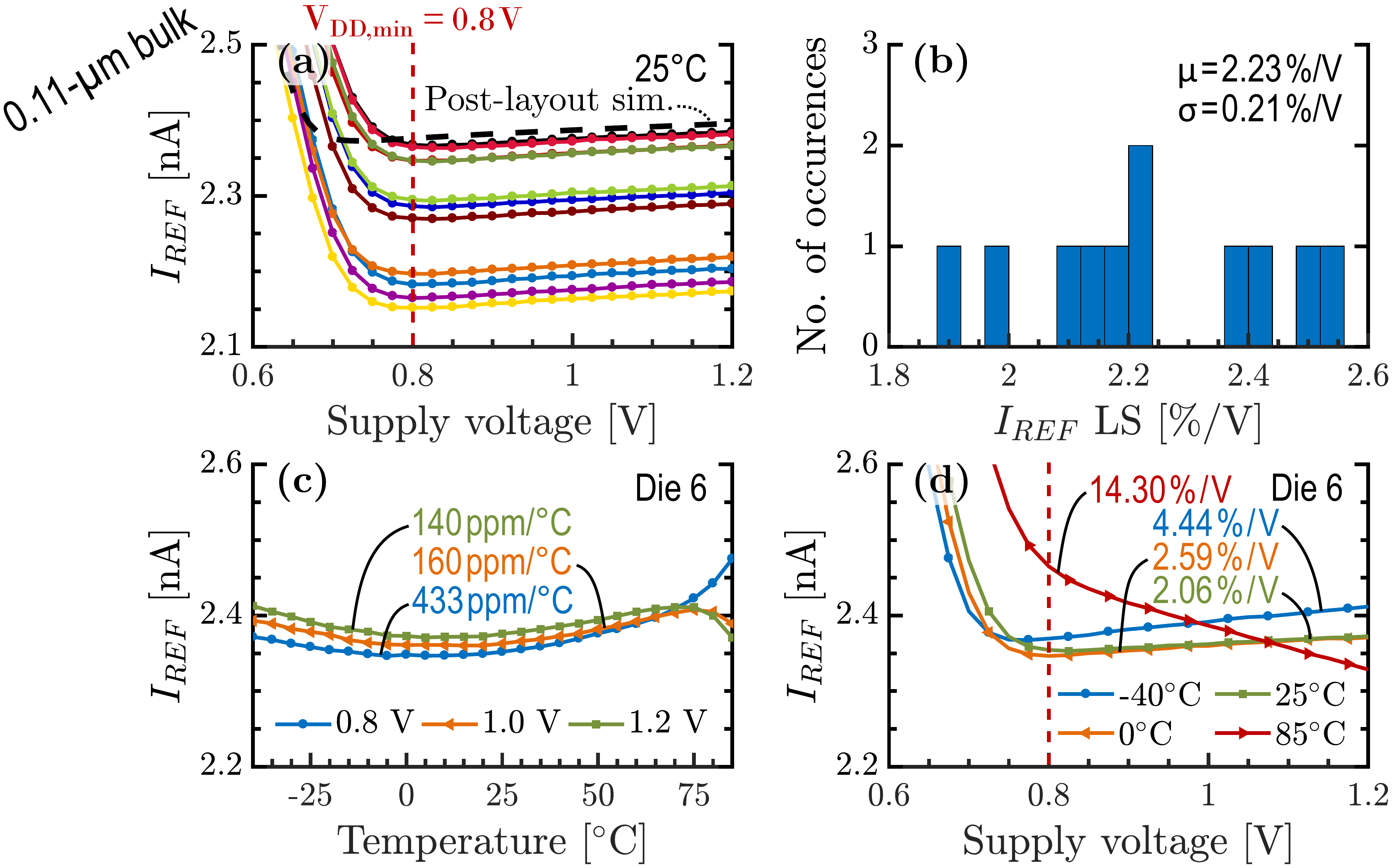}
	\caption{In UMC \mbox{0.11-$\mu$m} bulk and for the design with $I_{REF}$ TC calibration, (a) measured supply voltage dependence at 25$^\circ$C and (b) histogram of LS from 0.8 to 1.2~V. For die 6, measured (c) temperature dependence at different supply voltages and (d) supply voltage dependence at different temperatures.}
	\label{fig:20_meas_iref_vs_vdd_0p11um}
\end{figure}
\begin{figure}[!t]
	\centering
	\includegraphics[width=.5\textwidth]{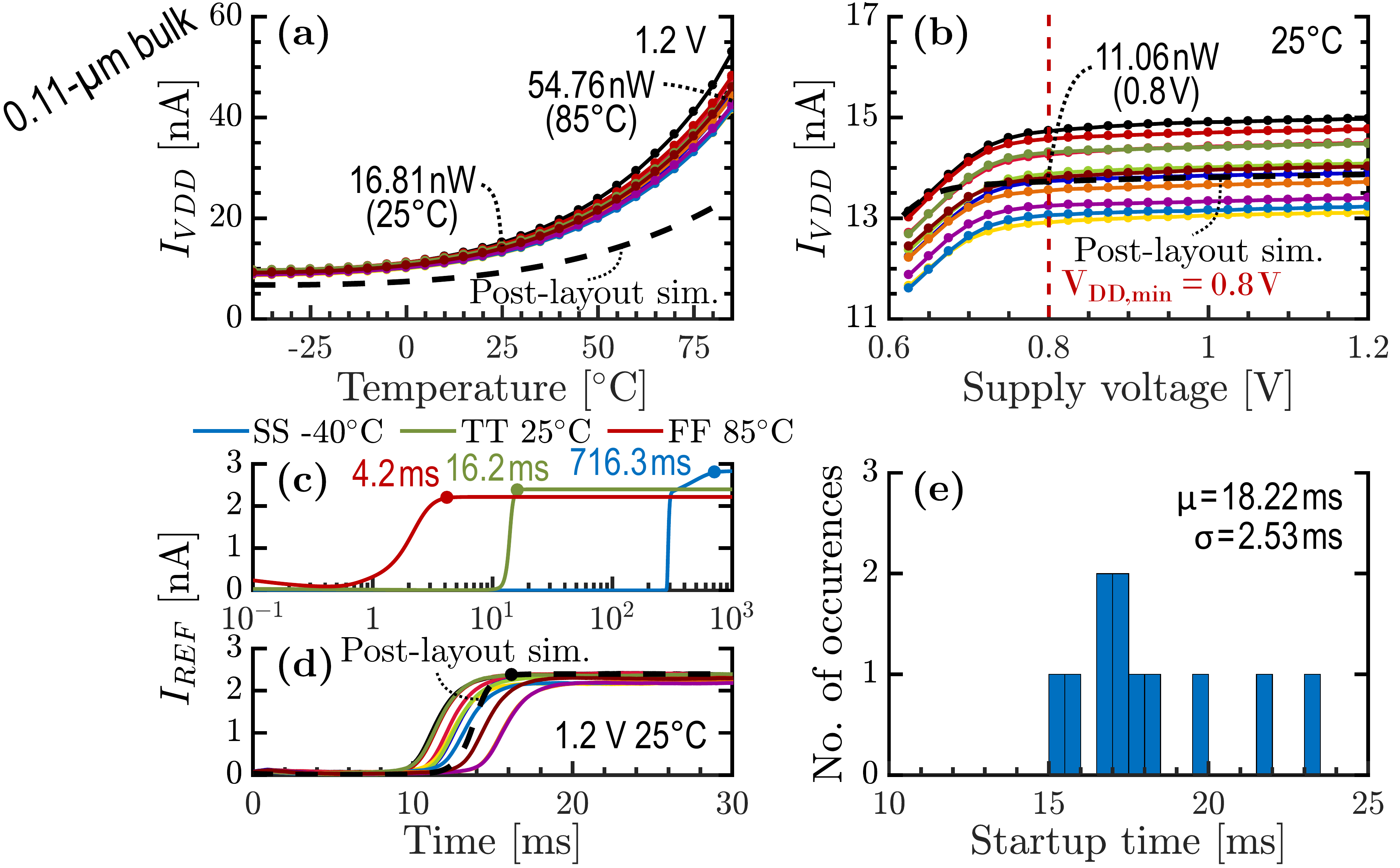}
	\caption{In UMC \mbox{0.11-$\mu$m} bulk, measured dependence of the supply current (a) to temperature at 1.2~V, and (b) to supply voltage at 25$^\circ$C. (c) Post-layout-simulated RCC startup waveforms in extreme corners, and (d) measured startup waveforms and (e) histogram of \mbox{99-$\%$} startup time, at 1.2~V 25$^\circ$C.}
	\label{fig:21_meas_ivdd_startup_0p11um}
\end{figure}
\begin{figure}[!t]
	\centering
	\includegraphics[width=.45\textwidth]{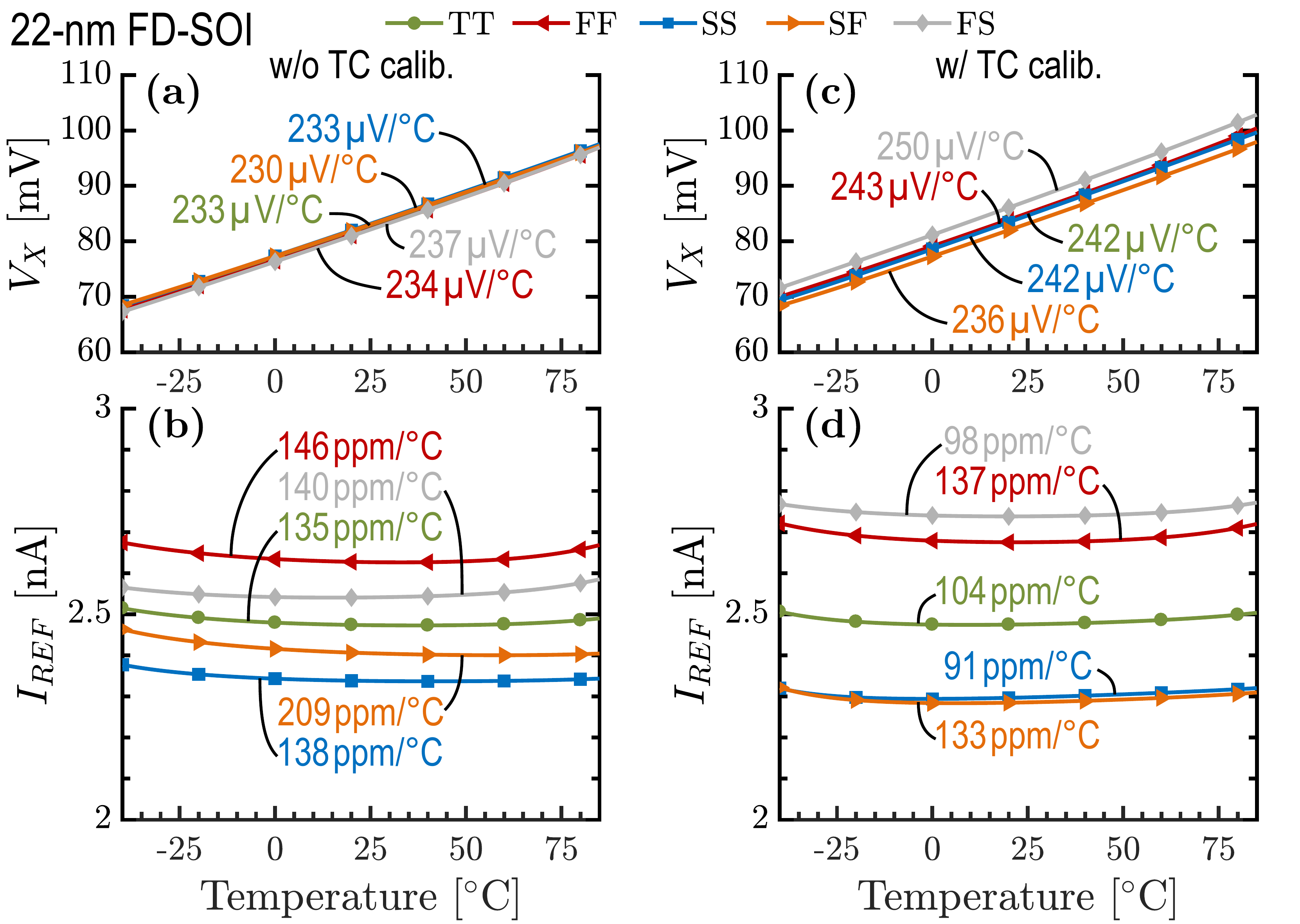}
	\caption{In GF \mbox{22-nm} \mbox{FD-SOI}, post-layout simulation of the temperature dependence of $V_X$ and $I_{REF}$, in all process corners and at 1.8~V, without [(a) and (b)] and with $I_{REF}$ TC calibration [(c) and (d)].}
	\label{fig:22_sim_iref_vs_T_22nm}
\end{figure}
\begin{figure}[!t]
	\centering
	\includegraphics[width=.45\textwidth]{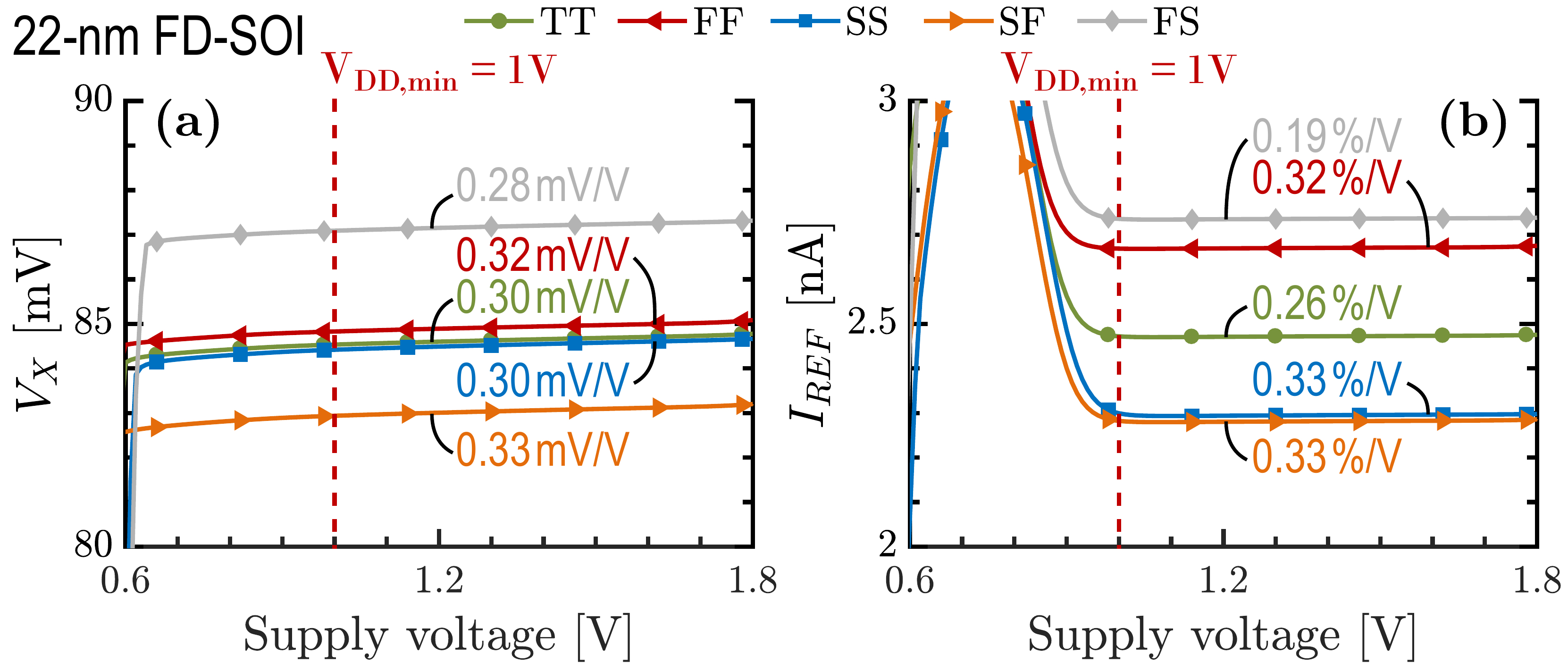}
	\caption{In GF \mbox{22-nm} \mbox{FD-SOI}, post-layout simulation of the supply voltage dependence of (a) $V_X$ and (b) $I_{REF}$ in all process corners and at 25$^\circ$C, with $I_{REF}$ TC calibration.}
	\label{fig:23_sim_iref_vs_vdd_22nm}
\end{figure}
We notice a drop of $I_{REF}$ at high temperature which is due to the leakage of the electrostatic discharge (ESD) protection diodes in the I/O pad connected to \mbox{SMU 1}. In Figs.~\ref{fig:19_meas_iref_vs_T_0p11um}(c) and (d), average values of 2.74 and 2.30~nA are obtained, with $(\sigma/\mu)$'s of 2.11 and 3.47~$\%$. This variability is above the simulated one for local mismatch, suggesting that the measured dies originate from different process batches. It should be noted that, if required by the applicative context, a calibration of $I_{REF}$ could readily be implemented using a binary-weighted current mirror. Finally, the calibration improves the TC from 530 to 176~ppm/$^\circ$C, and also reduces $\sigma$ from 95 to 38~ppm/$^\circ$C [Figs.~\ref{fig:19_meas_iref_vs_T_0p11um}(e) and (f)]. It should be noted that the 575-ppm/$^\circ$C outlier in Fig.~\ref{fig:19_meas_iref_vs_T_0p11um}(f), corresponding to the black curve in Fig.~\ref{fig:19_meas_iref_vs_T_0p11um}(b), is not accounted for in the mean. These results are obtained with a complete knowledge of the temperature profile between -40 and 85$^\circ$C, but a two-point calibration at -25 and 85$^\circ$C yields similar results, with a mean TC marginally increased to 182~ppm/$^\circ$C. Moving on to the supply voltage dependence, the details of the 11 dies are shown in Fig.~\ref{fig:20_meas_iref_vs_vdd_0p11um}(a), and an average LS of 2.23~$\%$/V is achieved in Fig.~\ref{fig:20_meas_iref_vs_vdd_0p11um}(b). This value is close to the 2.07~$\%$/V obtained in TT post-layout simulation, which is not surprising given the good agreement between simulation and measurement in Fig.~\ref{fig:20_meas_iref_vs_vdd_0p11um}(a). Then, Fig.~\ref{fig:20_meas_iref_vs_vdd_0p11um}(c) highlights that the $I_{REF}$ drop at high temperature transforms into an $I_{REF}$ surge as supply voltage decreases. This behavior arises from the fact that, at 85$^\circ$C, $I_{REF}$ decreases with $V_{DD}$ while it presents a normal supply voltage dependence at all other temperatures [Fig.~\ref{fig:20_meas_iref_vs_vdd_0p11um}(d)]. For a 0.8-V supply, the leakage of the ESD diodes between the pad and the supply voltages (core and I/O) dominates and increases the measured $I_{OUT}$, while at 1.2~V, the leakage of the diodes between the pad and ground is prevalent and decreases the measured $I_{OUT}$. Next, in Fig.~\ref{fig:21_meas_ivdd_startup_0p11um}(a), the supply current is around 10~nA at low temperature as it is dominated by the SCM which draws a current equal to $(N+1)I_{REF}$ with a current ratio $N=3$, while at high temperature, it scales exponentially as the power consumption of the 4T voltage reference is proportional to the $I_{DS}$ leakage. At 1.2~V, average power consumptions of 16.8 and 54.8~nW are reached at 25 and 85$^\circ$C, respectively. The steeper supply current increase in measurement compared to the TT simulation indicates that the dies might be from a fast nMOS process corner. Fig.~\ref{fig:21_meas_ivdd_startup_0p11um}(b) shows that the drawn current does not substantially increase with the supply voltage, and that a minimum average power consumption of 11.1 nW is achieved at 0.8~V. Finally, the \mbox{x-$\%$} startup time is computed as the time at which $I_{REF}$ remains within (100-x)~$\%$ of its steady-state value. The \mbox{99-$\%$} startup time has a simulated nominal value of 16.2~ms and a worst-case one of 716.3~ms in the SS -40$^\circ$C corner [Fig.~\ref{fig:21_meas_ivdd_startup_0p11um}(c)]. At 1.2~V and 25$^\circ$C, the measured startup time is 18.2~ms on average [Figs.~\ref{fig:21_meas_ivdd_startup_0p11um}(d) and (e)], and confirms the inutility of a startup circuit.
\begin{figure}[!t]
	\centering
	\includegraphics[width=.45\textwidth]{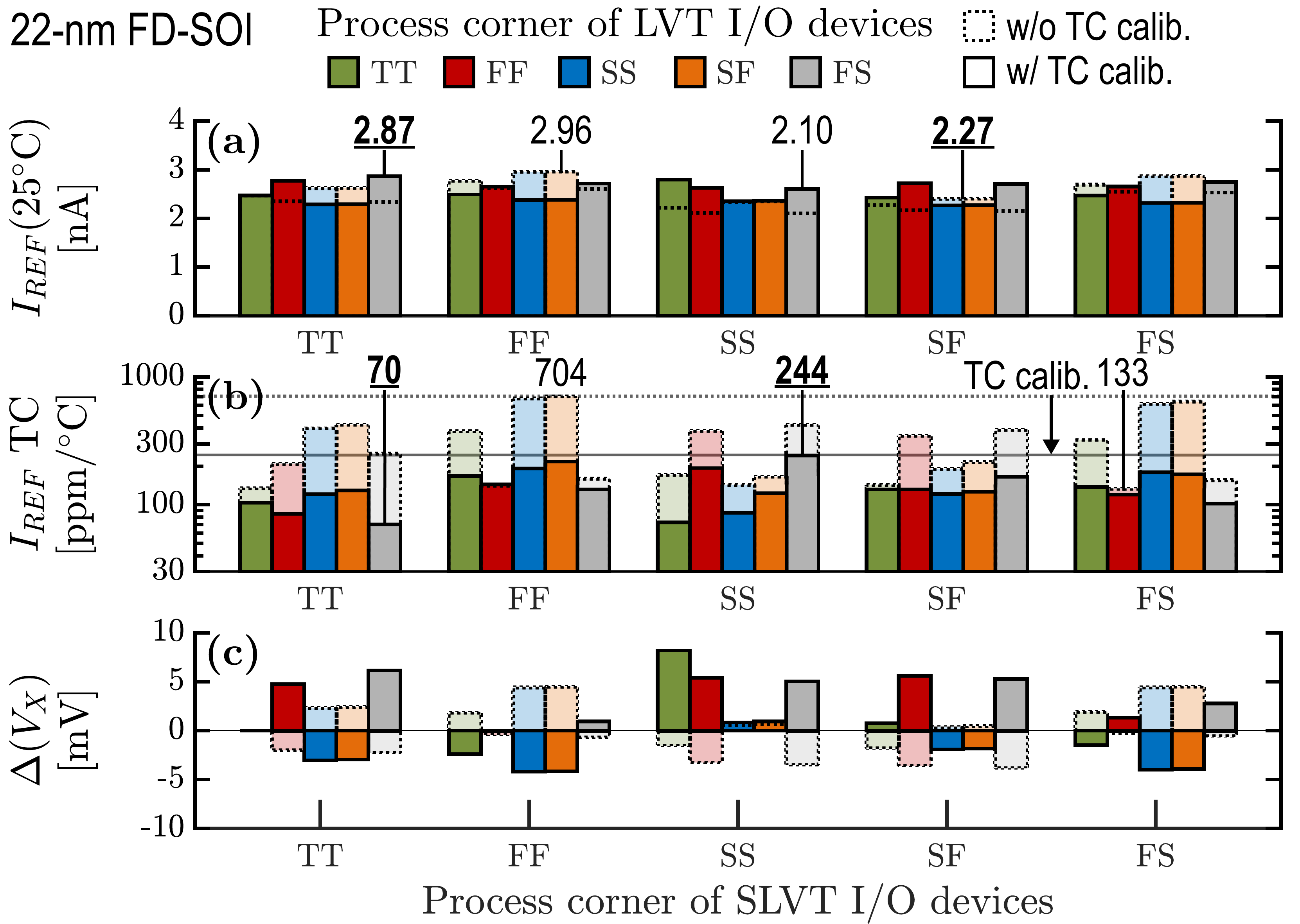}
	\caption{In GF \mbox{22-nm} \mbox{FD-SOI} and at 1.8~V, post-layout simulation of (a) $I_{REF}$ at 25$^\circ$C, (b) $I_{REF}$ TC from -40 to 85$^\circ$C, and (c) the change in PTAT slope of $V_X$ with respect to its nominal value, without and with TC calibration. Skewed process corners of SLVT and LVT I/O devices are considered.}
	\label{fig:24_sim_iref_vs_T_skewed_process_22nm}
\end{figure}
\begin{figure}[!t]
	\centering
	\includegraphics[width=.5\textwidth]{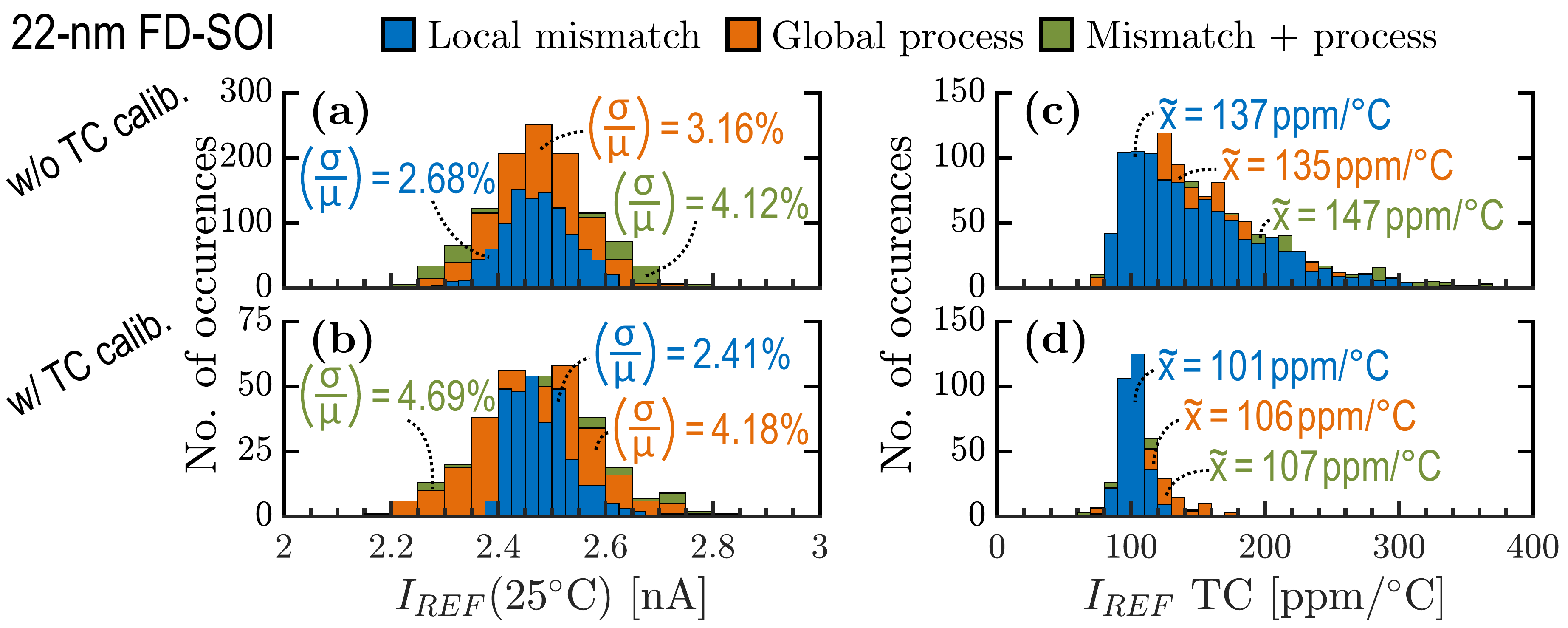}
	\caption{In GF \mbox{22-nm} \mbox{FD-SOI}, for 10$^3$ / 3$\times$10$^2$ post-layout MC simulations in TT at 1.8~V, histograms of $I_{REF}$ at 25$^\circ$C [(a) and (b)] and of $I_{REF}$ TC from -40 to 85$^\circ$C [(c) and (d)], without [(a) and (c)] and with $I_{REF}$ TC calibration [(b) and (d)]. $\tilde{x}$ denotes the median of a statistical distribution.}
	\label{fig:25_sim_iref_vs_T_mc_22nm}
\end{figure}

\subsection{Designs in 22-nm FD-SOI CMOS Technology}
\label{subsec:4D_designs_in_22-nm_FD-SOI_CMOS_technology}
\begin{figure}[!t]
	\centering
	\includegraphics[width=.5\textwidth]{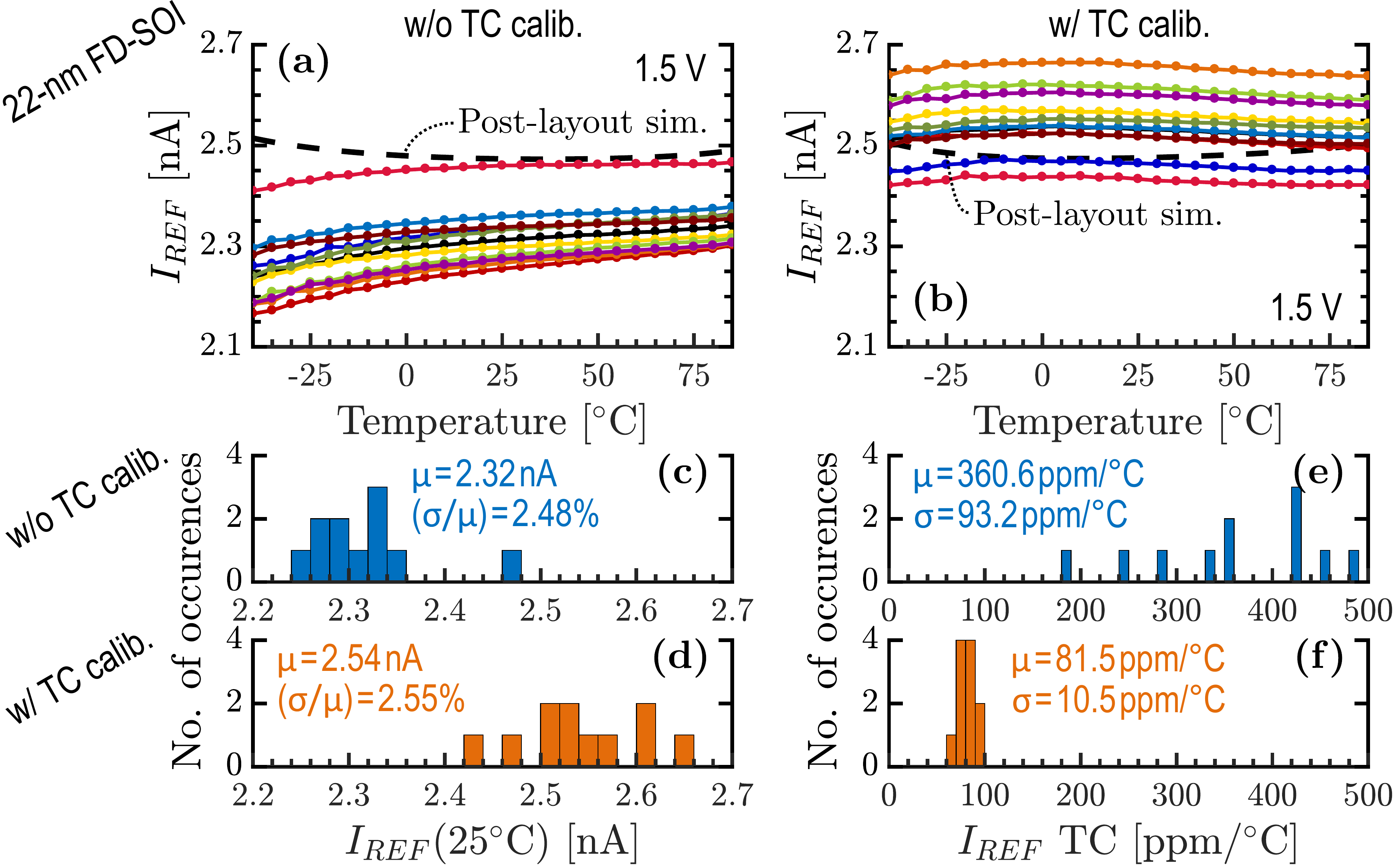}
	\caption{In GF \mbox{22-nm} \mbox{FD-SOI} and at 1.5~V, measured temperature dependence of $I_{REF}$ (a) without and (b) with $I_{REF}$ TC calibration. Measured histograms of $I_{REF}$ at 25$^\circ$C [(c) and (d)] and of $I_{REF}$ TC from -40 to 85$^\circ$C [(e) and (f)], without and with $I_{REF}$ TC calibration.}
	\label{fig:26_meas_iref_vs_T_22nm}
\end{figure}
\begin{figure}[!t]
	\centering
	\includegraphics[width=.5\textwidth]{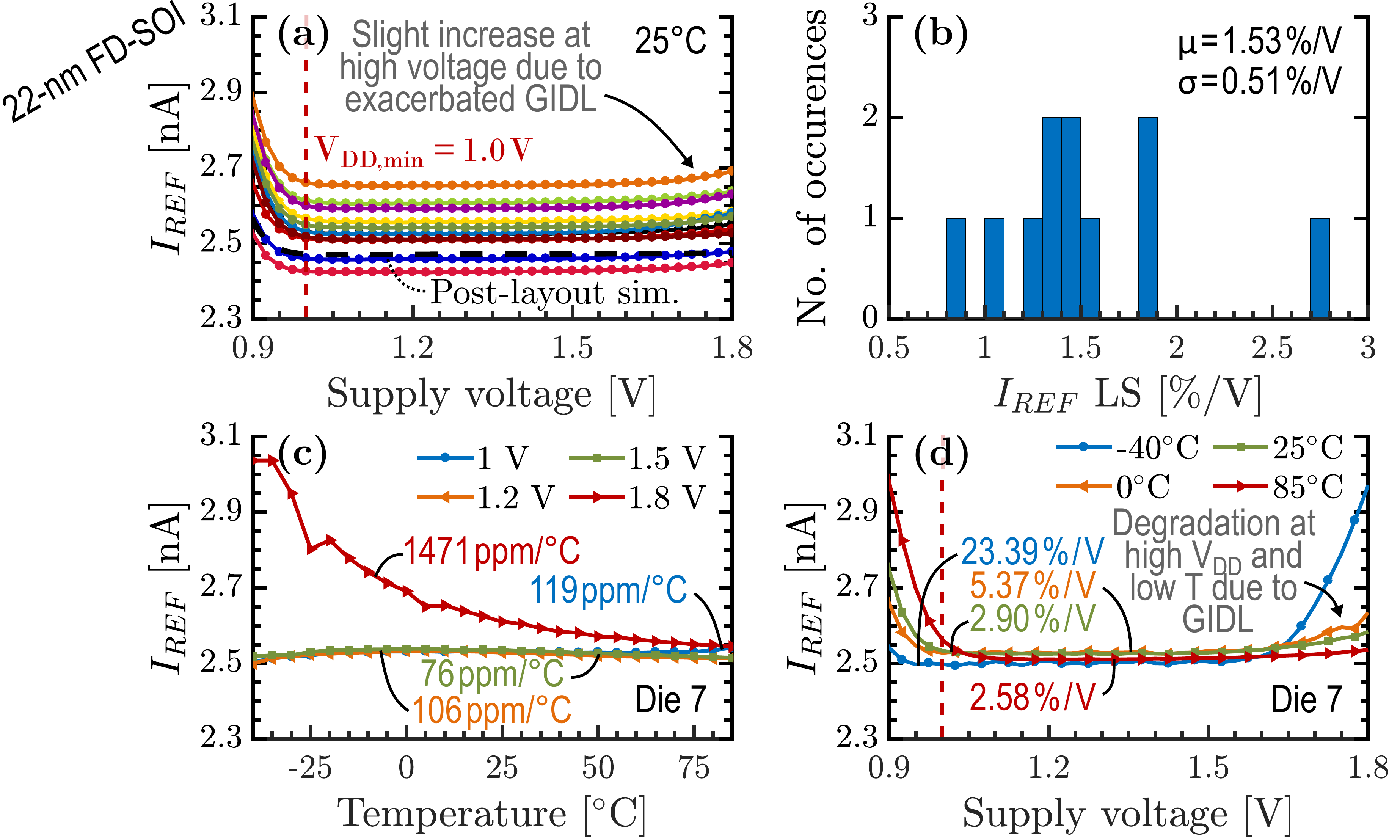}
	\caption{In GF \mbox{22-nm} \mbox{FD-SOI} and for the design with $I_{REF}$ TC calibration, (a) measured supply voltage dependence at 25$^\circ$C and (b) histogram of LS from 1 to 1.8~V. For die 7, measured (c) temperature dependence at different supply voltages and (d) supply voltage dependence at different temperatures.}
	\label{fig:27_meas_iref_vs_vdd_22nm}
\end{figure}
Similarly to the results in 0.11~$\mu$m, we first discuss the post-layout simulations results. Fig.~\ref{fig:22_sim_iref_vs_T_22nm}(a) shows that, without TC calibration, $\delta_{V_X}$ is less process-dependent than in 0.11~$\mu$m, and takes lower values between 230 and 237~$\mu$V/$^\circ$C. These two observations are linked to the subthreshold slope factor, which can be defined as \mbox{$n = 1 + C_d/C_{ox}$}. In bulk, $C_d$ is the depletion layer capacitance, while in FD-SOI, $C_d$ is the capacitance between the back-gate and the channel due to the buried oxide. This capacitance is process-invariant at first order, and smaller than the depletion layer capacitance, thus explaining our observations. The resulting $I_{REF}$ TC is comprised between 138 and 209~ppm/$^\circ$C [Fig.~\ref{fig:22_sim_iref_vs_T_22nm}(b)]. In 22~nm, the TC calibration changes $V_{off}$ [Fig.~\ref{fig:22_sim_iref_vs_T_22nm}(c)] and reduces $I_{REF}$ TC between 91 to 137~ppm/$^\circ$C [Fig.~\ref{fig:22_sim_iref_vs_T_22nm}(d)]. The necessity of the TC calibration might not be obvious from the analysis of conventional process corners, but will become clear when discussing skewed process corners herebelow. Besides, Fig.~\ref{fig:23_sim_iref_vs_vdd_22nm} shows that, for the reference with TC calibration, $V_X$ LS is around 0.28 to 0.32~mV/V [Fig.~\ref{fig:23_sim_iref_vs_vdd_22nm}(a)], which is more than 3$\times$ lower than in 0.11~$\mu$m thanks to the large intrinsic gain $(g_m/g_d)$ in FD-SOI \cite{Cathelin_2017}, thereby improving the LS, as it is usually proportional to $(g_d/g_m)$. This translates into an $I_{REF}$ LS between 0.19 and 0.33~$\%$/V from 1 to 1.8~V, with $V_{DD,\textrm{min}}$ = 1~V also limited by $V_G$ in the SCM. The larger $V_{DD,\textrm{min}}$ compared to 0.11~$\mu$m stems from a larger $V_G$, likely due to a different value of $I_{SQ}$ and/or $V_{T0}$ for transistors $M_{1-2}$ implementing the SCM. Next, we consider skewed process variations of SLVT and LVT I/O devices in Fig.~\ref{fig:24_sim_iref_vs_T_skewed_process_22nm}.
Contrary to the \mbox{0.11-$\mu$m} design, the TC calibration somewhat reduces $I_{REF}$ process variations from \mbox{+18.4}~$\%\:$/$\:$\mbox{-16}~$\%$ to \mbox{+14.8}~$\%\:$/$\:$\mbox{-9.2}~$\%$ [Fig.~\ref{fig:24_sim_iref_vs_T_skewed_process_22nm}(a)]. The max./min. values are however not attained in the same corners without and with TC calibration.
\begin{figure}[!t]
	\centering
	\includegraphics[width=.5\textwidth]{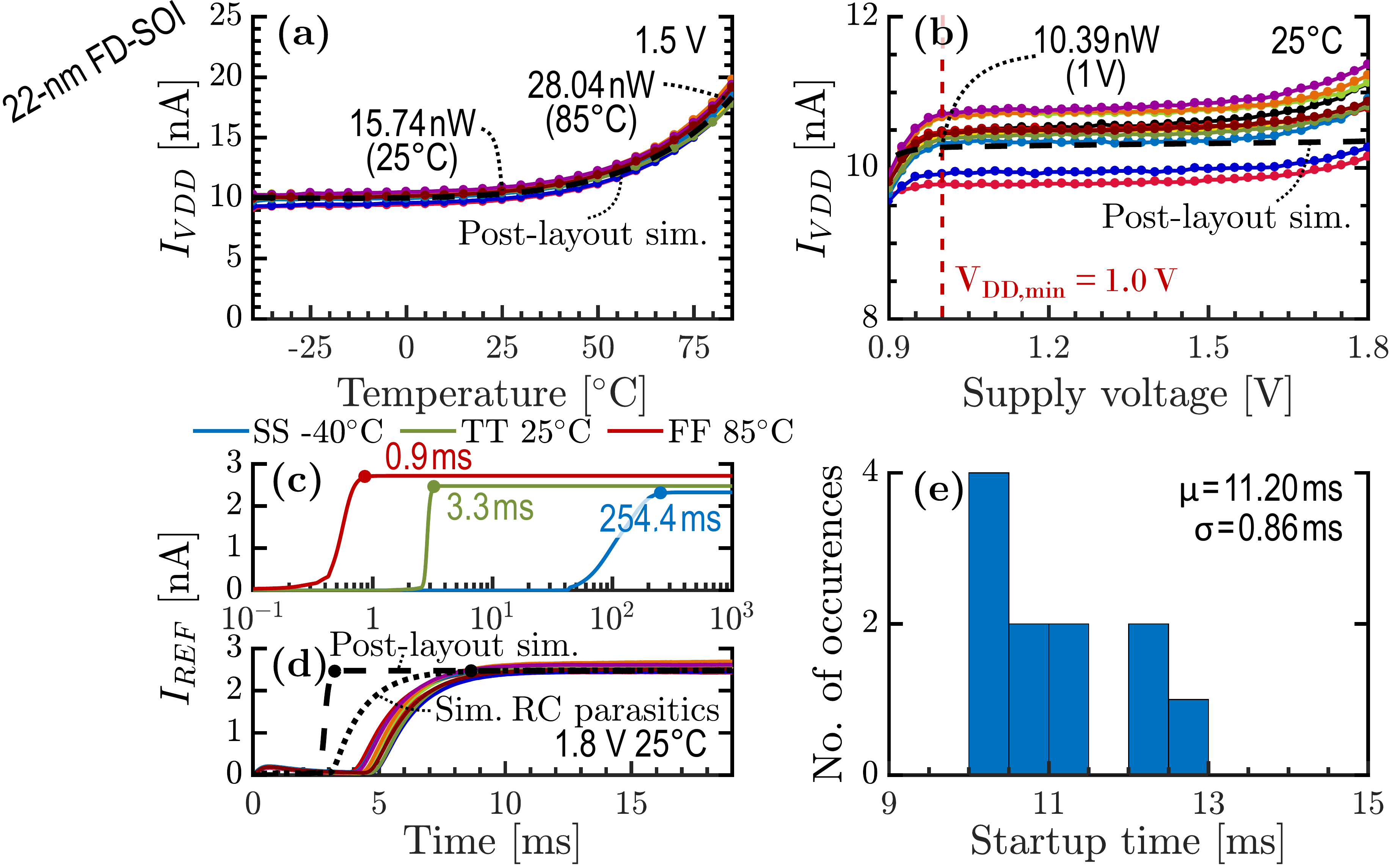}
	\caption{In GF \mbox{22-nm} \mbox{FD-SOI}, measured dependence of the supply current (a) to temperature at 1.5~V, and (b) to supply voltage at 25$^\circ$C. (c) Post-layout-simulated RCC startup waveforms in extreme corners, and (d) measured startup waveforms and (e) histogram of \mbox{99-$\%$} startup time, at 1.8~V 25$^\circ$C.}
	\label{fig:28_meas_ivdd_startup_22nm}
\end{figure}
\setlength{\tabcolsep}{2pt}
\begin{table*}[!t]
\centering
\caption{Comparison table of temperature-independent nA-range current references.}
\vspace{-0.15cm}
\label{table:soa_nanoamp_range}
\scalebox{.7}{%
\begin{threeparttable}
\begin{scriptsize}
\begin{tabular}[t]{l|ccccccccc|cccccccccc@{\hskip 5pt}|cc}
\toprule
Type of work & \multicolumn{9}{c}{\textbf{Simulations}} & \multicolumn{12}{c}{\textbf{Silicon measurements}} \\
\cmidrule(lr){2-10} \cmidrule(lr){11-22}
& Far & Cordova & Santamaria & Agarwal & Aminzadeh & Mahmoudi & Bruni & Huang & Yang & De Vita & Kayahan & Ji & Wang & Wang & Huang & Lee & Chang & Shetty & Lefebvre & \multicolumn{2}{c}{\textbf{Lefebvre}}\\
& \cite{Far_2015} & \cite{Cordova_2017} & \cite{Santamaria_2019} & \cite{Agarwal_2022} & \cite{Aminzadeh_2022} & \cite{Mahmoudi_2022} & \cite{Bruni_2023} & \cite{Huang_2023} & \cite{Yang_2023} & \cite{DeVita_2007} & \cite{Kayahan_2013} & \cite{Ji_2017} & \cite{Wang_2019_VLSI} & \cite{Wang_2019_TCAS} & \cite{Huang_2020} & \cite{Lee_2020} & \cite{Chang_2022} & \cite{Shetty_2022} & \cite{Lefebvre_2023} & \multicolumn{2}{c}{\textbf{This work}}\\
\midrule
Publication & ROPEC & ISCAS & ISCAS & TCAS-II & AEU &  & CAE & AEU & AEU & ISCAS & TCAS-I & ISSCC & VLSI-DAT & TCAS-I & TCAS-II & JSSC & JJAP & TCAS-I & JSSC & \multicolumn{2}{c}{JSSC} \\
Year & 2015 & 2017 & 2019 & 2022 & 2022 & 2022 & 2023 & 2023 & 2023 & 2007 & 2013 & 2017 & 2019 & 2019 & 2020 & 2020 & 2022 & 2022 & 2023 & \multicolumn{2}{c}{2024}\\
\cmidrule(lr){21-22}
Samples & N/A & N/A & N/A & N/A & N/A & N/A & N/A & N/A & N/A & 20 & 90 & 10 & 10 & 16 & 10 & 10 & 3 & 10 & 20 & 11 & 11\\
\midrule
Technology & 0.18$\mu$m & 0.18$\mu$m & 0.18$\mu$m & 0.18$\mu$m & 0.18$\mu$m & 0.13$\mu$m & 0.18$\mu$m & 0.18$\mu$m & 0.18$\mu$m & 0.35$\mu$m & 0.35$\mu$m & 0.18$\mu$m & 0.18$\mu$m & 0.18$\mu$m & 0.18$\mu$m & 0.18$\mu$m & 90nm & 0.13$\mu$m & 22nm & 0.11$\mu$m & 22nm\\
$I_{REF}$ [nA] & 14 & 10.9 & 2.7 & 5.6 & 6.7 & 6.6 & 6.3 & 8.9 & 1.96 & 9.1 & 25 & 6.7 & 6.5 & 9.8 & 11.6 & 1 & 1.3 & 1.9 & 1.25/0.9\tnote{$\star$} & 2.4/2.3\tnote{$\star$} & 2.5/2.5\tnote{$\star$}\\
Power [nW] & 150 & 30.5 & 26 & 9.5 & 51 & 3.7 & 3.3 & 0.05 & 9.2 & 109.7 & 28500 & 9.3 & 15.8 & 28 & 48.6 & 4.5/14 & 8.6 & 30 & 7.8/5.8\tnote{$\star$} & 16.7/16.8\tnote{$\star$} & 15.5/16.3\tnote{$\star$}\\
 & $@$1V & $@$0.9V & $@$2V & $@$0.55V & $@$1V & $@$0.4V & $@$0.6V & $@$0.8V & $@$0.55V & $@$3V & $@$5V & $@$N/A & $@$0.85V & $@$0.7V & $@$0.8V & $@$1.5V & $@$0.75V & $@$2.5V & $@$0.9V & $@$1.2V & $@$1.5V\\
Area [mm$^2$] & \textcolor{ECS-Blue}{\textbf{0.0102}} & \textcolor{ECS-Blue}{\textbf{0.01}} & \textcolor{ECS-Blue}{\textbf{0.0093}} & 0.032 & \textcolor{ECS-Red}{\textbf{0.46}} & \textcolor{ECS-Blue}{\textbf{0.0021}} & \textcolor{ECS-Blue}{\textbf{0.0018}} & \textcolor{ECS-Blue}{\textbf{0.008}} & \textcolor{ECS-Blue}{\textbf{0.0033}} & 0.035 & \textcolor{ECS-Blue}{\textbf{0.0053}} & \textcolor{ECS-Red}{\textbf{0.055}} & \textcolor{ECS-Red}{\textbf{0.062}} & \textcolor{ECS-Red}{\textbf{0.055}} & \textcolor{ECS-Red}{\textbf{0.054}} & \textcolor{ECS-Red}{\textbf{0.332}} & 0.0175 & 0.0163 & 0.0132 & \textcolor{ECS-Blue}{\textbf{0.0106}} & \textcolor{ECS-Blue}{\textbf{0.00255}}\\
\midrule
Supply range [V] & 1 -- 3.3 & 0.9 -- 1.8 & 2 -- 3.63\tnote{$\diamond$} & 0.55 -- 1.9 & 1.1 -- 1.8 & 0.4 -- 1.6 & 0.6 -- 1.8 & 0.8 -- 1.8 & 0.55 -- 1.8 & 1.5 -- 4 & N/A & 1.3 -- 1.8 & 0.85 -- 2 & 0.7 -- 1.2 & 0.8 -- 2 & 1.5 -- 2 & 0.75 -- 1.55 & 0.85 -- 2 & 0.9 -- 1.8 & 0.8 -- 1.2 & 1 -- 1.8\\
LS [$\%$/V] & \textcolor{ECS-Blue}{\textbf{0.1}} & \textcolor{ECS-Blue}{\textbf{0.54}} & \textcolor{ECS-Red}{\textbf{8.9}}\tnote{$\diamond$} & \textcolor{ECS-Blue}{\textbf{0.022}} & \textcolor{ECS-Blue}{\textbf{0.03}} & 2.7 & \textcolor{ECS-Red}{\textbf{12.1}} & 1.39 & \textcolor{ECS-Blue}{\textbf{0.2}} & \textcolor{ECS-Blue}{\textbf{0.57}} & \textcolor{ECS-Red}{\textbf{150}} & 1.16 & 4.15 & \textcolor{ECS-Blue}{\textbf{0.6}} & 1.08 & 1.4 & \textcolor{ECS-Blue}{\textbf{0.15}} & 4 & \textcolor{ECS-Blue}{\textbf{0.26}}/\textcolor{ECS-Blue}{\textbf{0.39}}\tnote{$\star$} & 2.07/2.23\tnote{$\star$} & \textcolor{ECS-Blue}{\textbf{0.26}}/1.53\tnote{$\star$}\\
\midrule
Temp. range [$^\circ$C] & 0 -- 70 & -20 -- 120 & -40 -- 125 & -30 -- 70 & -40 -- 120 & -40 -- 120 & -40 -- 120 & 0 -- 125 & 0 -- 100 & 0 -- 80 & 0 -- 80 & 0 -- 110 & -10 -- 100 & -40 -- 125 & -40 -- 120 & -20 -- 80 & 0 -- 120 & -40 -- 120 & -40 -- 85 & -40 -- 85 & -40 -- 85\\
TC [ppm/$^\circ$C] & \textcolor{ECS-Blue}{\textbf{20}} & \textcolor{ECS-Blue}{\textbf{108}} & 309 & 256 & \textcolor{ECS-Blue}{\textbf{40.3}} & 308 & 219 & \textcolor{ECS-Blue}{\textbf{139}} & \textcolor{ECS-Blue}{\textbf{96.8}} & \textcolor{ECS-Blue}{\textbf{44}} & \textcolor{ECS-Blue}{\textbf{128}}/250\tnote{$\star$} & \textcolor{ECS-Red}{\textbf{680}}/283\tnote{$\dagger$} & \textcolor{ECS-Blue}{\textbf{157}} & \textcolor{ECS-Blue}{\textbf{150}} & \textcolor{ECS-Blue}{\textbf{169}} & 289/265\tnote{$\triangleright$} & \textcolor{ECS-Blue}{\textbf{53}}/394\tnote{$\star$} & 530/\textcolor{ECS-Red}{\textbf{822}}\tnote{$\star$} & \textcolor{ECS-Blue}{\textbf{203}}/565\tnote{$\star$} & 290/\textcolor{ECS-Blue}{\textbf{176}}\tnote{$\star$} & \textcolor{ECS-Blue}{\textbf{101}}/\textcolor{ECS-Blue}{\textbf{82}}\tnote{$\star$}\\
TC type & Typ. & $\mu$ & $\mu$ & $\mu$ & $\mu$ & $\mu$ & Typ. & $\mu$ & $\mu$ & $\mu$ & N/A & $\mu$ & $\mu$ & $\mu$ & $\mu$ & $\mu$ & $\mu$ & $\mu$ & $\mu$ & $\mu$ & $\mu$\\
\midrule
$I_{REF}$ var. & \multirow{2}{*}{N/A} & 15.8 / & \multirow{2}{*}{N/A} & +55.4 / & \multirow{2}{*}{N/A} & \multirow{2}{*}{N/A} & +129.1 / & +2.8 / & \multirow{2}{*}{8.7} & \multirow{4}{*}{2.16} & \multirow{2}{*}{8/1.22\tnote{$\star$}} & \multirow{2}{*}{N/A} & \multirow{2}{*}{N/A} & +11.7 / & +17.6 / & \multirow{2}{*}{N/A} & \multirow{4}{*}{21.1} & \multirow{2}{*}{N/A} & +9.9 / & +34.8 / & +15.9 /\\
(process) [$\%$] & & 11.6\tnote{$\dagger$} & & -28.5\tnote{$\diamond$} &  &  & -61.8\tnote{$\diamond$} & -13.9\tnote{$\diamond$} &  &  &  &  &  & -8.7\tnote{$\diamond$} & -10.3\tnote{$\diamond$} &  &  &  & -9.5 & -25.9 & -8.3\\
$I_{REF}$ var. & \multirow{2}{*}{5.8} & \multirow{2}{*}{N/A} & \multirow{2}{*}{\textcolor{ECS-Red}{\textbf{20.3}}} & \multirow{2}{*}{\textcolor{ECS-Red}{\textbf{10.4}}} & \multirow{2}{*}{\textcolor{ECS-Blue}{\textbf{0.7}}} & \multirow{2}{*}{\textcolor{ECS-Red}{\textbf{6.1}}} & \multirow{2}{*}{N/A} & \multirow{2}{*}{2.6} & \multirow{2}{*}{1.7} &  & \multirow{2}{*}{1.4} & \multirow{2}{*}{4.07/1.19\tnote{$\star$}} & \multirow{2}{*}{3.33} & \multirow{2}{*}{1.6} & \multirow{2}{*}{4.3} & \multirow{2}{*}{1.26/\textcolor{ECS-Blue}{\textbf{0.25}}\tnote{$\dagger$}} & & \multirow{2}{*}{\textcolor{ECS-Red}{\textbf{15.6}}} & \multirow{2}{*}{\textcolor{ECS-Red}{\textbf{6.39}}/\textcolor{ECS-Red}{\textbf{9.20}}\tnote{$\star$}} & \multirow{2}{*}{1.74/3.47\tnote{$\star$}} & \multirow{2}{*}{2.41/2.55\tnote{$\star$}}\\
(mismatch) [$\%$] &  &  &  &  &  &  &  &  &  &  &  &  &  &  &  &  &  &  &  &  &\\
\midrule
Trimming & No & Yes & No & No & No & Yes & No & Yes & No & No & No & Yes & No & Yes & Yes & Yes & Yes & Yes & No & Yes & Yes\\
Spec. components & No & \textcolor{ECS-Red}{\textbf{ZVT}} & No & Res. & Res., \textcolor{ECS-Red}{\textbf{BJT}} & No & No & No & No & No & No & Res., \textcolor{ECS-Red}{\textbf{BJT}} & No & Res. & Res. & No & No & Res. & No & No & No\\
\midrule
FoM & \multirow{2}{*}{\textcolor{ECS-Blue}{\textbf{0.0029}}} & \multirow{2}{*}{\textcolor{ECS-Blue}{\textbf{0.0077}}} & \multirow{2}{*}{0.0174} & \multirow{2}{*}{0.0819} & \multirow{2}{*}{0.1159} & \multirow{2}{*}{\textcolor{ECS-Blue}{\textbf{0.0040}}} & \multirow{2}{*}{\textcolor{ECS-Blue}{\textbf{0.0025}}} & \multirow{2}{*}{\textcolor{ECS-Blue}{\textbf{0.0089}}} & \multirow{2}{*}{\textcolor{ECS-Blue}{\textbf{0.0032}}} & \multirow{2}{*}{0.0193} & \multirow{2}{*}{0.0166} & \multirow{2}{*}{0.1415} & \multirow{2}{*}{0.0887} & \multirow{2}{*}{0.0499} & \multirow{2}{*}{0.0570} & \multirow{2}{*}{0.9595} & \multirow{2}{*}{0.0575} & \multirow{2}{*}{0.0835} & \multirow{2}{*}{0.0597} & \multirow{2}{*}{0.0149} & \multirow{2}{*}{\textcolor{ECS-Blue}{\textbf{0.0017}}}\\
$[$ppm/$^\circ$C$^2 \times$mm$^2]$ &  &  &  &  &  &  &  &  &  &  &  &  &  &  &  &  &  &  &  &  &\\
\bottomrule
\end{tabular}
\end{scriptsize}
\begin{footnotesize}
\begin{tablenotes}
	\item[$\star$] Simulated and measured values.
	\item[$\dagger$] Before and after trimming.
	\item[$\diamond$] Estimated from figures.
	\item[$\triangleright$] For 25 and 2.5 minutes between two calibrations.
\end{tablenotes}
\end{footnotesize}
\end{threeparttable}%
}
\end{table*}
$I_{REF}$ TC is reduced from the 133-to-704-ppm/$^\circ$C range down to the 70-to-204-ppm/$^\circ$C one, pointing out the importance of this calibration to maintain performance across process corners [Fig.~\ref{fig:24_sim_iref_vs_T_skewed_process_22nm}(b)]. We notice that the effect of the TC calibration is to slightly increase $V_{off}$ in fast LVT nMOS corners, and to decrease it in slow LVT nMOS corners [Fig.~\ref{fig:24_sim_iref_vs_T_skewed_process_22nm}(c)]. Regarding $I_{REF}$ variability at 25$^\circ$C, similarly to the 0.11-$\mu$m references, the impact of local mismatch is more limited than that of global process variations, but with values that are nonetheless closer from each other, with a \mbox{2.68-$\%$} and \mbox{3.16-$\%$} $(\sigma/\mu)$'s for mismatch and process variations without TC calibration [Fig.~\ref{fig:25_sim_iref_vs_T_mc_22nm}(a)]. With TC calibration, the impact of mismatch is reduced to a \mbox{2.41-$\%$} $(\sigma/\mu)$, while the impact of process is a bit more significant due to the TC calibration itself, with a value of 4.18~$\%$ [Fig.~\ref{fig:25_sim_iref_vs_T_mc_22nm}(b)]. The median TC improves by roughly 40~ppm/$^\circ$C when the TC is calibrated, and the tail of the distribution is also reduced, with the 99th percentile for combined variations diminished from 333 to 239~ppm/$^\circ$C [Figs.~\ref{fig:25_sim_iref_vs_T_mc_22nm}(c) and (d)].\\
\indent Continuing with the measurement results, Figs.~\ref{fig:26_meas_iref_vs_T_22nm}(a) and (b) depict the temperature dependence without and with TC calibration for a supply voltage of 1.5~V, instead of 1.8~V in post-layout simulation results.
\begin{figure}[!t]
	\centering
	\includegraphics[width=.45\textwidth]{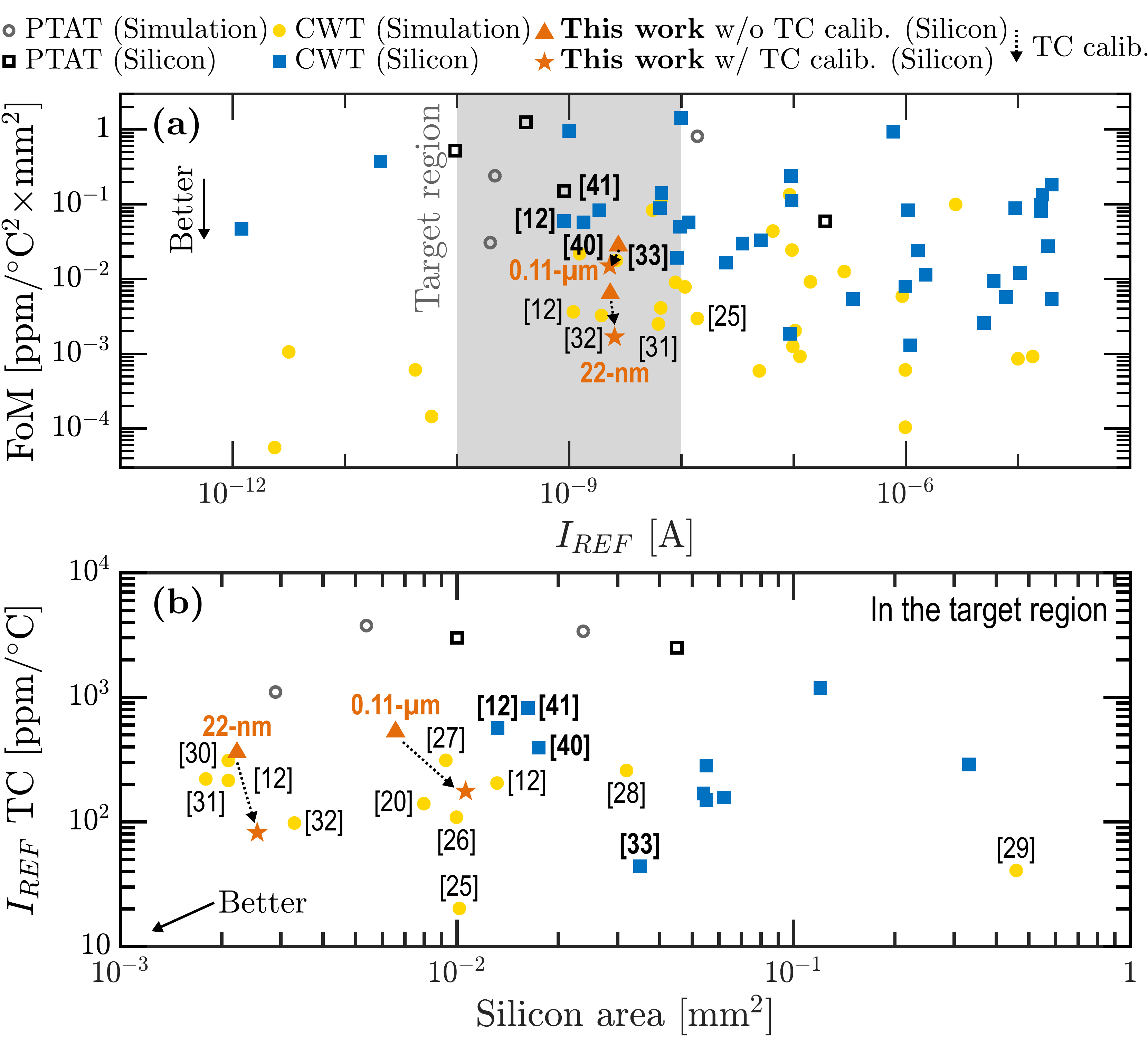}
	\caption{(a) Figure of merit (\ref{eq:FoM}) combining temperature dependence and silicon area as a function of $I_{REF}$, and (b) tradeoff between TC and area in the nA target region, based on the state of the art of current references.}
	\label{fig:29_comparison_to_soa}
\end{figure}
This choice is related to nonidealities in the 4T voltage reference and will be explained later in this paragraph. Figs.~\ref{fig:26_meas_iref_vs_T_22nm}(c) and (d) present 2.32- and \mbox{2.54-nA} average values of $I_{REF}$ without and with TC calibration, with $(\sigma/\mu)$'s of 2.48 and 2.55~$\%$. These results are in line with post-layout simulations and suggest that the measured dies come from the same process batch, as the variability has the same magnitude as the one due to local mismatch in simulation. The TC calibration mechanism cuts the TC from 361 to 82~ppm/$^\circ$C, with a sharp reduction of $\sigma$ from 93 to 11~ppm/$^\circ$C. In addition, it is interesting to note that all measured dies present a TC lower than 100~ppm/$^\circ$C. Similarly to the \mbox{0.11-$\mu$m} design, these results correspond to a calibration based on the complete temperature profile between -40 and 85$^\circ$C, but a two-point calibration at -40 and 80$^\circ$C gives comparable results and only slightly increases the mean TC to 82.7~ppm/$^\circ$C. Furthermore, Fig.~\ref{fig:27_meas_iref_vs_vdd_22nm}(a) reveals a modest rise of $I_{REF}$ at high supply voltage, due to the exacerbated GIDL at high $V_{DS}$ in the zero-$V_{GS}$ transistors used as current sources in the 4T voltage reference. This results in an average LS of 1.53~$\%$/V larger than the \mbox{0.26-$\%$/V} simulated value [Fig.~\ref{fig:27_meas_iref_vs_vdd_22nm}(b)]. Besides, Fig.~\ref{fig:27_meas_iref_vs_vdd_22nm}(c) illustrates that the TC is consistent across the supply voltage range, except at 1.8~V, for which $I_{REF}$ soars due to the exacerbated GIDL of the zero-$V_{GS}$ transistors, made worse by the increased $V_{T0}$ at low temperature. This interpretation is supported by the measurements in Fig.~\ref{fig:27_meas_iref_vs_vdd_22nm}(d), in which the distortion at high supply voltage is aggravated by the temperature reduction, and hints that nMOS devices might be in a slow process corner. Nevertheless, this problem could be solved by placing diode transistors in series with the zero-$V_{GS}$ transistors $M_{7Vi}$ to reduce their $V_{DS}$, without potentially impacting $V_{DD,\textrm{min}}$ as it is currently limited by voltage $V_G$ in the SCM. Then, the supply current temperature dependence [Fig.~\ref{fig:28_meas_ivdd_startup_22nm}(a)] is similar to the \mbox{0.11-$\mu$m} design, with a prevalence of the SCM current equal to 4$I_{REF} \approx$ 10~nA at low temperature, and an increased share due to the 4T voltage reference as temperature rises. Average power consumptions of 15.8 and 28.0~nW are respectively attained at 25 and 85$^\circ$C. Moreover, the supply voltage dependence of $I_{VDD}$ is similar to that of $I_{REF}$ [Fig.~\ref{fig:28_meas_ivdd_startup_22nm}(b)], as the power consumption at 25$^\circ$C is dominated by the SCM, and a minimum power consumption of 10.4~nW is obtained for $V_{DD,\textrm{min}}$ = 1~V. At last, the simulated \mbox{99-$\%$} startup time is 3.3~ms in typical conditions and 254.4~ms in the worst case [Fig.~\ref{fig:28_meas_ivdd_startup_22nm}(c)], while the measured value is 11.2~ms on average [Figs.~\ref{fig:28_meas_ivdd_startup_22nm}(d) and (e)]. On the one hand, this difference is explained by an nMOS process corner slightly slower than the typical one, which causes the shift to the right in Fig.~\ref{fig:28_meas_ivdd_startup_22nm}(d) and corroborates the observations related to the temperature and supply voltage dependences. On the other hand, it is due to the RC time constant caused by the series resistance $R_S$ and the \mbox{25-pF} PCB parasitic capacitance, as illustrated in Fig.~\ref{fig:28_meas_ivdd_startup_22nm}(d). These results again prove that the proposed reference does not require a startup circuit.

\section{Comparison to the State of the Art}
\label{sec:5_comparison_to_the_state_of_the_art}
In this section, we compare our work to the state of the art of simulated and fabricated nA-range CWT current references. Table~\ref{table:soa_nanoamp_range} summarizes the performance of each reference, while Fig.~\ref{fig:29_comparison_to_soa} provides a graphical representation of the TC and silicon area across different current levels, using a new figure of merit combining these two characteristics that we propose.
\begin{equation}
	\textrm{FoM} = \frac{\textrm{TC}}{(T_{\mathrm{max}} - T_{\mathrm{min}})} \times \textrm{Area}\:[\textrm{ppm}/^\circ\textrm{C}^2 \times \textrm{mm}^2] \label{eq:FoM}
\end{equation}
This FoM shares some similarities with the one in \cite{CamposDeOliveira_2018}, introduced in the context of voltage references and later applied to current references \cite{Shetty_2022}. However, the proposed FoM must be minimized, and removes the square on the temperature range as we do not believe it is relevant to put so much emphasis on this characteristic. Power consumption normalized to the reference current could also be integrated to this FoM as
\begin{equation}
	\textrm{FoM}_2 = \textrm{FoM} \times \frac{\textrm{Power}}{(\textrm{1 V}\times I_{REF})}\:[\textrm{ppm}/^\circ\textrm{C}^2 \times \textrm{mm}^2]\textrm{.} \label{eq:FoM_2}
\end{equation}
\indent First, the proposed designs uses more scaled technologies than the state of the art, predominantly featuring 0.18~$\mu$m. Nonetheless, we did not normalize the silicon area to the technology node for several reasons:
\begin{enumerate}
	\item Analog circuits do not benefit from technology scaling as much as digital ones, as they have to cope with exacerbated analog nonidealities in advanced nodes \cite{Lewyn_2009};
	\item An alternative scaling, whose idea is to maintain the same variability $\sigma_{V_T}$ in any technology, could rely on Pelgrom's mismatch parameter $A_{V_T}$. Using Pelgrom's law with a constant $\sigma_{V_T}$, we observe that silicon area is proportional to $A_{V_T}^2$. Based on values of $A_{V_T}$ of 4.5~mV$\times\mu$m and 3~mV$\times\mu$m in 0.18 and 0.11~$\mu$m \cite{Lewyn_2009}, and 2.4~mV$\times\mu$m in 28~nm \cite{Cathelin_2017} (close to 22~nm used in this work), the scaling factor with respect to 0.18~$\mu$m would be 2.3$\times$ for 0.11~$\mu$m and 3.5$\times$ for 28~nm, as opposed to 2.7$\times$ and 41.3$\times$ for a conventional scaling based on the feature size. The latter thus disproportionately penalizes advanced technologies. Although this alternative scaling based on $A_{V_T}$ might seem attractive, it is not conventionally done in the literature and difficult to apply in a fair and rigorous way, as a circuit consists of several transistors of different types and lengths and $A_{V_T}$ is precisely impacted by these two parameters;
	\item The area of some current references \cite{Ji_2017, Wang_2019_TCAS, Huang_2020} is dominated by the resistor used as $V$-to-$I$ converter, for which the scaling factor to be used is not obvious;
	 \item In 22~nm, we used I/O devices whose \mbox{0.15-$\mu$m} $L_{min}$ is comparable to 0.18~$\mu$m and whose $A_{V_T}$ exceeds that of core devices due to the increased oxide thickness \cite{Mezzomo_2011}.
\end{enumerate}
\noindent Besides, Fig.~\ref{fig:29_comparison_to_soa}(a) highlights that calibrating $I_{REF}$ TC improves the FoM of the proposed \mbox{0.11-$\mu$m} and \mbox{22-nm} designs by 1.9$\times$ and 3.8$\times$, respectively, by reducing the TC while slightly increasing the silicon area [Fig.~\ref{fig:29_comparison_to_soa}(b)]. Therefore, in what follows, we will focus on the designs with $I_{REF}$ TC calibration as they present a better FoM value.\\
\indent Next, compared to fabricated references in the literature, we obtain a 1.3$\times\:$/$\:$11.4$\times$ FoM reduction for the \mbox{0.11-$\mu$m} and \mbox{22-nm} designs compared to the closest measured competitor \cite{DeVita_2007}, achieving a \mbox{44-ppm/$^\circ$C} TC within a \mbox{0.035-mm$^2$} area. For both designs, the FoM improvement compared to \cite{DeVita_2007} stems from the lower limit of the temperature range, decreased from 0 to -40$^\circ$C, and the area reduction, as \cite{DeVita_2007} relies on a somewhat-complex $\beta$-multiplier variant. Furthermore, \cite{Wang_2019_VLSI, Wang_2019_TCAS, Huang_2020} feature a TC around 150~ppm/$^\circ$C that is competitive with the proposed designs, but exhibit a larger area above 0.05~mm$^2$, due to the use of resistors \cite{Wang_2019_TCAS, Huang_2020} or massive transistors in deep subthreshold \cite{Wang_2019_VLSI} as $V$-to-$I$ converters. Finally, \cite{Lefebvre_2023, Chang_2022, Shetty_2022} occupy a silicon area between 0.01 and 0.02~mm$^2$ which is close to the proposed designs, as \cite{Lefebvre_2023} relies on an SCM biased by a modified $\beta$-multiplier, \cite{Chang_2022} on a gate-leakage transistor biased by a 6T voltage reference, and \cite{Shetty_2022} on the weighted sum of PTAT and CTAT currents generated by BJT-based $\beta$-multipliers. Despite the area efficiency of these techniques, the $I_{REF}$ TC is worse than in the proposed designs as these architectures do not embed any TC calibration circuit, whose addition would inevitably increase the area they occupy.

\section{Conclusion}
\label{sec:6_conclusion}
In this work, we demonstrated a nA-range CWT current reference based on an SCM biased by a PTAT voltage with a CWT offset. This bias voltage is generated by a 4T ULP voltage reference which is the key innovation of this work as (i) it can integrate two types of simple and area-efficient $I_{REF}$ TC calibration circuits, which are key to maintain performance across process corners and to obtain a measured TC comparable with post-layout simulations, (ii) it makes it possible to tune the CWT offset based on transistor sizes, allowing to reliably attain a good performance in terms of LS and variability, in any technology and within a limited area, and (iii) it does not require any startup circuit. Moreover, the proposed current reference architecture can also be used to generate a current with a specific temperature dependence by employing a different sizing or by exploiting the TC calibration circuit. Then, we presented a methodology for sizing the proposed reference based on the ACM model and supported by post-layout simulations. Lastly, we validated the proposed reference based on designs without and with $I_{REF}$ TC calibration, fabricated in \mbox{0.11-$\mu$m} bulk and \mbox{22-nm} \mbox{FD-SOI} technologies, to prove that the body$\:$/$\:$back-gate effect can indeed be leveraged in these two technology types. The designs with $I_{REF}$ TC calibration achieve a \mbox{2.3-nA} current with a \mbox{176-ppm/$^\circ$C} TC and a \mbox{2.23-$\%$/V} LS in 0.11~$\mu$m, and a \mbox{2.5-nA} current with a \mbox{82-ppm/$^\circ$C} TC and a \mbox{1.53-$\%$/V LS} in 22~nm. They demonstrate the high level of performance achieved by the proposed reference, as well as its portability to different technologies. In addition, the simplicity of the proposed architecture renders its sizing and implementation quite straightforward. Further work should however focus on reducing the power consumption of the 4T voltage reference at high temperature.


%


\vspace{-0.25cm}
\section*{Acknowledgments}
The authors would like to thank Pierre G\'{e}rard for the measurement testbench, El\'{e}onore Masarweh for the microphotograph, and ECS group members for their proofreading.

\ifCLASSOPTIONcaptionsoff
  \newpage
\fi



%
\bibliographystyle{IEEEtran}
\bibliography{Lefebvre_JSSC_2023_MANTIS_CERBERUS_current_ref}
\vspace{-1cm}

%

\begin{IEEEbiography}[{\includegraphics[width=1in,height=1.25in,clip,keepaspectratio]{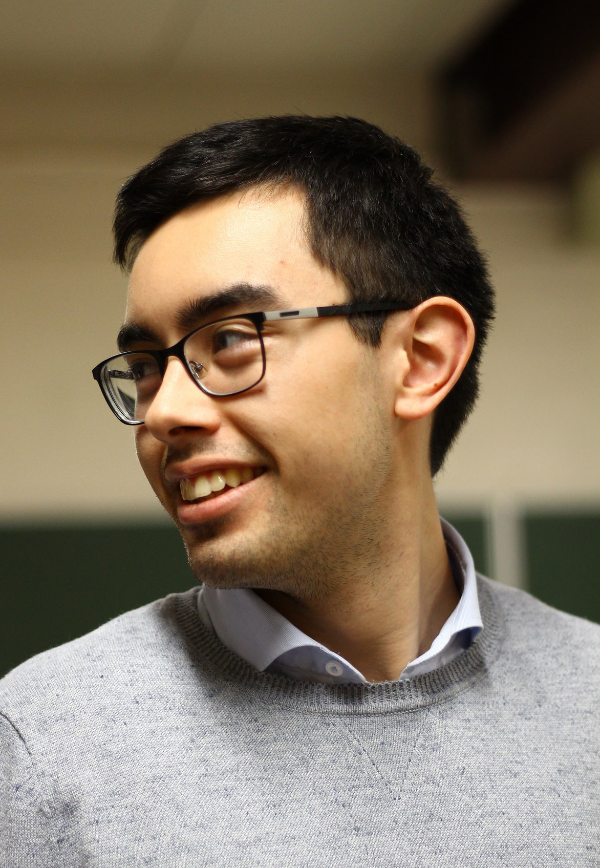}}]{Martin Lefebvre} (Graduate Student Member, IEEE) received the M.Sc. degree (summa cum laude) in Electromechanical Engineering and the Ph.D. degree from the Universit\'e catholique de Louvain (UCLouvain), Louvain-la-Neuve, Belgium, in 2017 and 2024, respectively. His Ph.D. thesis, focusing on area-efficient and temperature-independent current references for the Internet of Things, was supervised by Prof. David Bol. His current research interests include hardware-aware machine learning algorithms, low-power mixed-signal vision chips for embedded image processing, and ultra-low-power current reference architectures. He serves as a reviewer for various IEEE conferences and journals, including J. Solid-State Circuits, Trans. on Circuits and Syst., Trans. on VLSI Syst., and Int. Symp. on Circuits and Syst. (ISCAS).
\end{IEEEbiography}
\vspace{-1cm}

\begin{IEEEbiography}[{\includegraphics[width=1in,height=1.25in,clip,keepaspectratio]{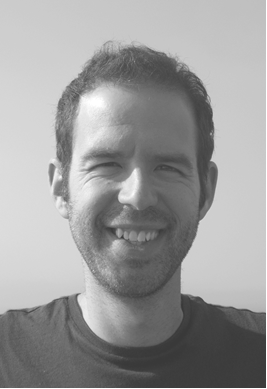}}]{David Bol} (Senior Member, IEEE) is an Associate Professor at UCLouvain. He received the Ph.D. degree in Engineering Science from UCLouvain in 2008 in the field of ultra-low-power digital nanoelectronics. In 2005, he was a visiting Ph.D. student at the CNM, Sevilla, and in 2009, a post-doctoral researcher at intoPIX, Louvain-la-Neuve. In 2010, he was a visiting post-doctoral researcher at the UC Berkeley Lab for Manufacturing and Sustainability, Berkeley. In 2015, he participated to the creation of e-peas semiconductors spin-off company. Prof. Bol leads the Electronic Circuits and Systems (ECS) group focused on ultra-low-power design of integrated circuits for environmental and biomedical IoT applications including computing, power management, sensing and wireless communications. He is actively engaged in a social-ecological transition in the field of ICT research with a post-growth approach. Prof. Bol has authored more than 150 papers and conference contributions and holds three delivered patents. He (co-)received four Best Paper/Poster/Design Awards in IEEE conferences (ICCD 2008, SOI Conf. 2008, FTFC 2014, ISCAS 2020) and supervised the Ph.D. thesis of Charlotte Frenkel who received the 2021 Nokia Bell Scientific Award and the 2021 IBM Innovation Award for her Ph.D. He serves as a reviewer for various IEEE journals and conferences and presented several keynotes in international conferences. On the private side, Prof. Bol pioneered the parental leave for male professors in his faculty, to spend time connecting to nature with his family. 
\end{IEEEbiography}




\end{document}